\documentclass[]{aastex631}

\shorttitle{Age of metal poor stars}
\shortauthors{Plotnikova et al.}

\graphicspath{{./}{figures/}}

\begin{document}

\title{Very metal-poor stars in the solar vicinity: age determination}

\author[0000-0002-7504-0950]{Anastasiia Plotnikova}
\affiliation{Dipartimento di Fisica e Astronomia, Universit\'a di Padova,I-35122, Padova, Italy}

\author[0000-0002-0155-9434]{Giovanni Carraro}
\affiliation{Dipartimento di Fisica e Astronomia, Universit\'a di Padova,I-35122, Padova, Italy}

\author[0000-0001-6205-1493]{Sandro Villanova}
\affiliation{Departamento de Astronomia, Casilla 160-C, Universidad de Concepci{\'o}n, Concepci{\'o}n. Chile}

\author[0000-0001-7939-5348]{Sergio Ortolani}
\affiliation{Dipartimento di Fisica e Astronomia, Universit\'a di Padova,I-35122, Padova, Italy}

\begin{abstract} 
The ages of the oldest and most metal-poor stars in the Milky Way bear important information on the age of the Universe and its standard model.
We analyze a sample of 28 extremely metal-poor field stars in the solar vicinity culled from the literature and carefully determine their ages. To this aim, we critically make use of Gaia data to derive their distances and associated uncertainties. Particular attention has been paid to the estimate of the reddening and its effect on the derivation of stellar ages. We employed different reddenings and super-impose isochrones from different sources in the stars color-magnitude diagram built up with different photometric systems. We highlight subtle metallicity effects when using the Johnson photometry for low metallicity stars and finally adopt Gaia photometry.
An automatic fitting method is devised to assign ages to each individual star taking into account the uncertainties in the input parameters. 
The mean age of the sample turns out to be $13.9 \pm 0.5$ Gyr using Padova isochrones, and $13.7 \pm 0.4$ Gyr using BASTI isochrones. We found also a group of very metal-poor stars ($\left[\frac{Fe}{H}\right]$: -2.7 -- -2.0 dex) with relatively young ages, in the range 8 -- 10 Gyr.

\end{abstract}


\section{Introduction} \label{sec:intro}

Stars with extremely low metal abundance are of particular astrophysical and cosmological interest because they probe very early times in the evolution of the Universe and its Galactic components. Through the investigation of the age and chemical composition, we can obtain important constraints on the evolution of the Milky Way, set up a lower limit to the age of the Universe (\cite{Bond2013}, \cite{VandenBerg2014}), and understand the chemical properties of the first Population III supernovae in the nascent Milky Way (\cite{Frebel2015}).

Over the years, several spectroscopic campaigns have been conducted to study the chemical composition of very metal-poor stars (\cite{Christlieb_2004}, \cite{Cayrel_2004}, \cite{Barklem2005}, \cite{Schlaufman_2014}, \cite{Limberg_2021}, etc.). The most recent investigations of the age of metal-poor stars (\cite{Bonaca_2020}, \cite{Carter_2021}) showed that they are on average old ($12.0 \pm 1.5$ Gyr for turnoff stars). However, both authors discovered the presence of very metal-poor stars with relatively young ages 8 - 10 Gyr (3 stars in \cite{Bonaca_2020}, 1 star in \cite{Carter_2021}).

The current best estimate of the age of the Universe is $13.77 \pm 0.06$ Gyr, based on the latest WMAP derivation (\cite{Bennett2013}), and it is in excellent agreement with observations of the cosmic microwave background (CMB) using the Planck satellite (\cite{Ade_2014}). Recent simulations (e.g., \cite{Ritter_2012}, \cite{Safranek-Shrader_2014}) suggest that the oldest Population II stars probably formed $\sim$ 0.2 -- 0.3 Gyr after the Big Bang, depending on how quickly the gas from the first (Population III) supernovae was able to cool down and condense, as well as on the relevance and impact of the Population III stellar feedback. Precise ages for the oldest and most metal-poor stars can date the onset of star formation (e.g., \cite{Bromm_2004}) following the Big Bang. Since the oldest stars must be younger than the Universe, precise ages provide a strong test of the consistency between cosmological and stellar physics.

Moreover, the derived age together with high precision spectroscopic measurements of metallicity can help us to reconstruct the age-metallicity relation (AMR) - one of the main observational constraints for any Milky Way formation and evolution model.  The first AMR in the solar neighborhood was obtained by \cite{Twarog_1980} and it shows an average decrease of the metallicity with the age. \cite{Twarog_1980} concluded that the AMR can be used to estimate the star formation rate via comparison with theoretical models. That means that the AMR is an important tracer of the star formation history of galaxies. An extension of the AMR to the very metal-poor tail will help to understand the early stages of star formation in the Milky Way and in the Universe.

On the other hand, the wide distribution in metallicity in the solar neighborhood, at any given age, suggests that stars have been moving away from their birth locations over time (radial migration) (\cite{Grenon_1972}, \cite{Grenon_1989}). The main mechanisms of radial migration are: transient spiral modes mostly at the co-rotation resonance (\cite{Sellwood_2002}), and non-axisymmetric perturbations to the potential such as bars or galaxy interactions (\cite{Bird_2012}, \cite{Roskar_2008}, \cite{Quillen_2009}).

A recent intriguing finding, described in \cite{Feuillet_2018}, is that the youngest stars are not the most metal-rich as previously expected from pure gas enrichment processes. There are two possible reasons for such an occurrence: 1) two modes of star formation, where part of the metal-rich gas is diluted to form stars that are subsequently more metal-poor, or 2) migration of the old, metal-rich stars from the inner to the outer regions of the Galaxy. That means that non-linear patterns in AMR are correlated with the specific events in star formation and dynamical evolution of the Galaxy.

In this study, ages and chemical compositions of a sample of very metal-poor stars are investigated. Our main goal is to compare them with the age of the Universe ($13.77 \pm 0.06$ Gyr) based on data on the CMB, baryon acoustic oscillations, and Hubble constant (\cite{Bennett2013}). We also investigate the chemical composition of these stars to retrieve information about the first Population III supernovae, the chemical composition of the nascent Milky Way, and the AMR behavior in the very metal-poor regime.

\section{Data}
As the main target of our research, we used metal-poor stars from the Hamburg vs. ESO R-process Enhanced Star (HERES) survey, spectroscopically studied by \cite{Barklem2005}. Their snapshot spectra cover a wavelength range of 3760 - 4980 \AA~  and have an average signal-to-noise ratio of S/N $\sim$ 54 per pixel over the entire spectral range. A $2''$ slit is employed giving a minimum resolving power of $R \approx$ 20000. From the "snapshot" spectra the elemental abundances of moderate precision (absolute rms errors of order 0.25 dex, relative rms errors of order 0.15 dex) have been obtained for 22 elements: C, Mg, Al, Ca, Sc, Ti, V, Cr, Mn, Fe, Co, Ni, Zn, Sr, Y, Zr, Ba, La, Ce, Nd, Sm, and Eu.

We chose this particular data set because of its careful and detailed chemical analysis. It is indeed a unique data set under this point of view. Besides, all stars in the data set under investigation are very metal-poor and this gives us the real opportunity to study the very beginning of the formation and evolution of our Galaxy. Additionally, we wanted to test the cosmological age of the Universe and very metal-poor stars are the best candidates for this purpose.

The final sample of stars analysed in \cite{Barklem2005} contains 253 stars built up with the following criteria:
\begin{itemize}
    \item [-] No strong molecular carbon features in spectra.
    \item [-] Metallicity cut-off: $\rm[Fe/H] < -1.5$
    \item [-] Temperature cut-off: $T_{eff} > 4200$ K
    \item [-] No spectroscopic binaries or rotators
\end{itemize}

\begin{figure}
\centering
\includegraphics[scale=0.2]{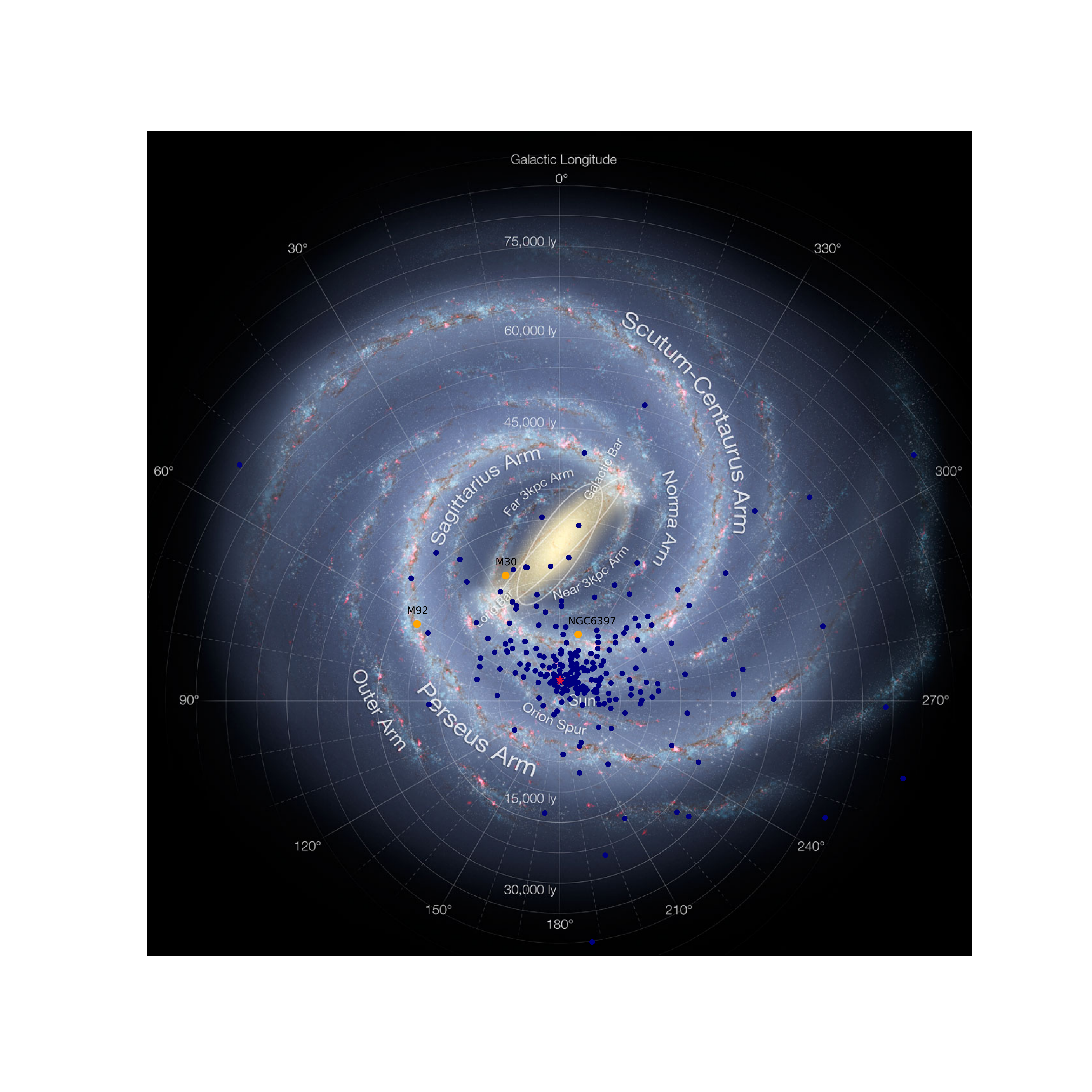}
\caption{Location of 253 very metal-poor stars (blue) in Galactic plane. Yellow dotes are metal-poor globular clusters: NGC 6397, M 30, M 92. Red stars are stars: HD 84937, HD 132475, and HD 140283 (\cite{VandenBerg2014})}
\label{fig:Location in galactic plane}
\end{figure}

We complemented spectroscopic data with Gaia (G, G$_{BP}$, G$_{RP}$) and Johnson (B, V, I) photometry obtained from the Gaia archive\footnote{\url{https://gea.esac.esa.int/archive/}} and SIMBAD catalog\footnote{\url{http://simbad.u-strasbg.fr/simbad/sim-fbasic}}.
The distribution in the galactic plane is shown in Fig.\ref{fig:Location in galactic plane}. Out of these 253 stars only 28 were TO stars for which we could measure good ages (see Section \ref{results}). Coordinates, photometry, interstellar absorption, distance, temperature and chemical abundances of these 28 stars are presented in Tab.\ref{tab:data}. See Sections \ref{Reddening} and \ref{Distance determination} for interstellar absorption and distance determination. The main characteristics of the data set:

\begin{itemize}
    \item [-] Location in the space: Galactic Halo $|b|>20^o$.
    \item [-] Most of the stars have magnitudes in the range $12 < G < 17$.
    \item [-] Distance (estimated using Gaia Data Release 3 parallaxes) in the range $0 < d < 30$ kpc.
    \item [-] Metallicity in the range $-3.8 < [Fe/H] < 1.5$ dex.
    \item [-] All stars are $\alpha$-element enhanced.
    \item [-] All targets are single stars. Spectroscopic binaries were already removed by \cite{Barklem2005}. We checked, however, \cite{El-Badry_2021} for spatially resolved binaries. No overlap was found.
\end{itemize}

\begin{figure}
\centering
\includegraphics[scale=0.5]{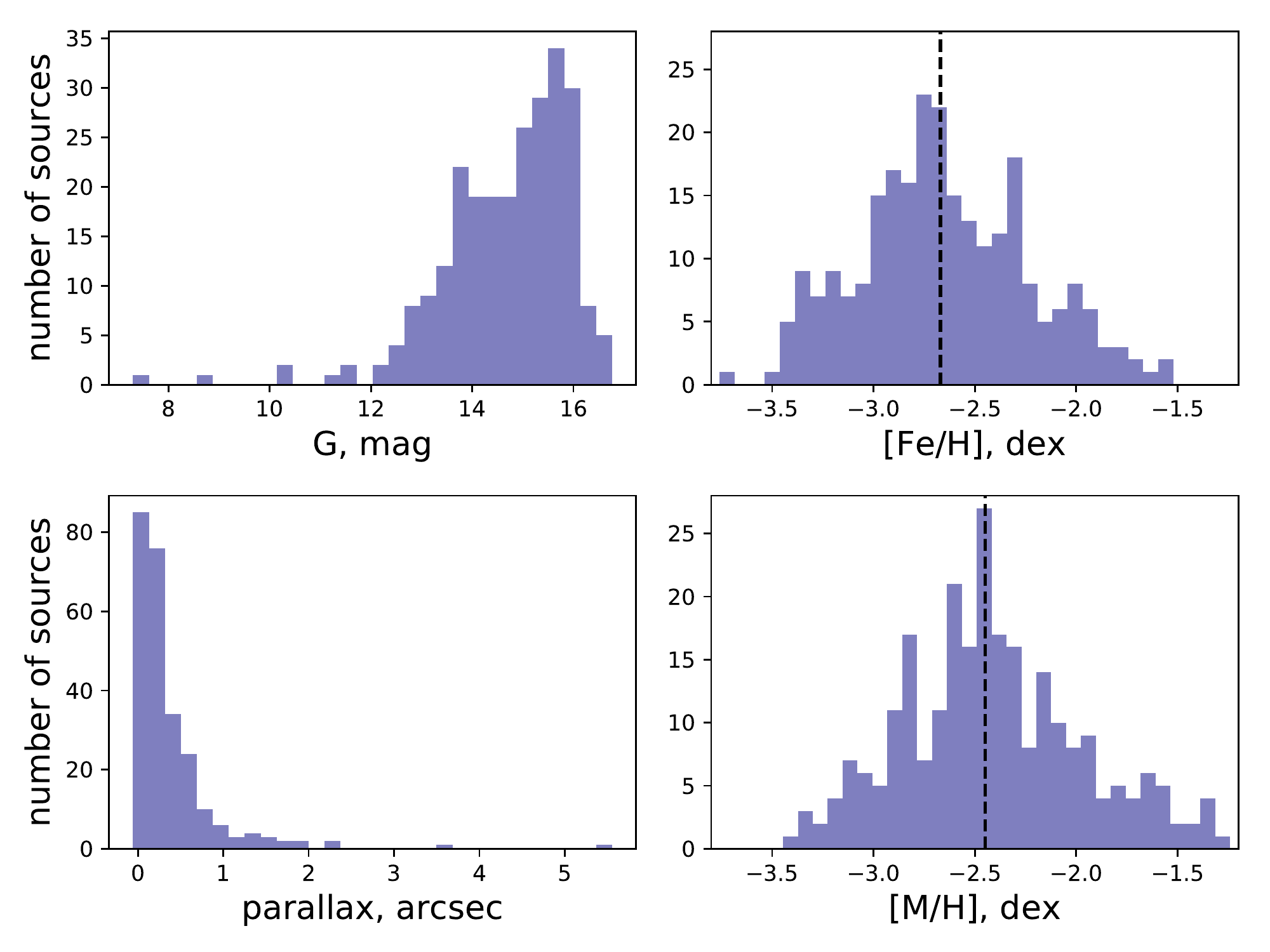}
\caption{Distribution of very metal-poor stars in G-band (Gaia DR3) - \textit{top left}, in parallax Gaia DR3 - \textit{bottom left}, in total metallicity \textit{top right}, in metallicity - \textit{bottom right}, vertical dashed line is the metallicity mean.}
\label{fig:Data distribution}
\end{figure}

\begin{rotatetable}
\begin{deluxetable*}{lrrlllllllrlll}
\tablecaption{Parameters for the stars analized in this study.\label{tab:data}}
\tablewidth{700pt}
\tabletypesize{\scriptsize}
\tablehead{ID & RA & DEC & B & V & I & G & G$_{BP}$ & G$_{RP}$ & A$_{V}$ & d & T$_{eff}$ & $\rm[Fe/H]$ & $\rm[C/Fe]$ \\
& J2000 & J2000 & mag & mag & mag & mag & mag & mag & mag & pc & K & dex & dex\\}

\startdata
HE\_0023-4825 & 00 25 50.31 & -48 08 27.01 & 14.298 & 13.830 & 13.161 & 13.667 & 13.945 & 13.218 & 0.032 & 1164 & 5816 & \nodata 2.06 & 0.31\\
HE\_0109-3711 & 01 11 38.40 & -36 55 17.11 & 16.668 & 16.290 & 15.691 & 16.160 & 16.400 & 15.757 & 0.025 & 3662 & 6156 & \nodata 1.91 & 0.31\\
HE\_0231-4016 & 02 33 44.39 & -40 03 42.72 & 16.494 & 16.090 & 15.475 & 15.939 & 16.185 & 15.532 & 0.036 & 2634 & 5972 & \nodata 2.08 & 1.36\\
HE\_0340-3430 & 03 42 04.76 & -34 20 50.10 & \nodata & 14.783 & 14.189 & 14.657 & 14.910 & 14.238 & 0.031 & 1780 & 5914 & \nodata 1.95 & 0.06\\
HE\_0430-4404 & 04 31 38.10 & -43 57 48.71 & \nodata & 15.724 & 15.166 & 15.629 & 15.868 & 15.236 & 0.035 & 1569 & 6214 & \nodata 2.07 & 1.44\\
HE\_0447-4858 & 04 49 01.00 & -48 53 36.15 & 16.687 & 16.254 & 15.624 & 16.140 & 16.398 & 15.727 & 0.038 & 3261 & 5995 & \nodata 1.69 & 0.04\\
HE\_0501-5139 & 05 02 48.21 & -51 35 36.30 & 16.573 & 16.094 & 15.475 & 15.980 & 16.250 & 15.543 & 0.059 & 4539 & 5861 & \nodata 2.38 & 0.40\\
HE\_0519-5525 & 05 19 59.15 & -55 22 41.81 & 15.570 & 15.034 & 14.284 & 14.862 & 15.171 & 14.374 & 0.066 & 2376 & 5580 & \nodata 2.52 & 0.29\\
HE\_0534-4615 & 05 35 52.94 & -46 13 35.97 & \nodata & 15.056 & 14.317 & 14.896 & 15.231 & 14.382 & 0.062 & 2266 & 5506 & \nodata 2.01 & 0.13\\
HE\_0938+0114 & 09 40 43.20 & +01 00 29.51 & \nodata & \nodata  & \nodata & 10.344 & 10.560 & 9.964 & 0.034 & 180 & 6777 & \nodata 2.51 & 0.65\\
HE\_1052-2548 & 10 55 20.53 & -26 04 48.03 & 13.492 & 13.188 & 12.541 & 13.115 & 13.364 & 12.699 & 0.127 & 675 & 6534 & \nodata 2.29 & 0.51\\
HE\_1105+0027 & 11 07 49.50 & +00 11 38.34 & 16.038 & 15.646 & 15.018 & 15.594 & 15.852 & 15.166 & 0.005 & 3162 & 6132 & \nodata 2.42 & 2.00\\
HE\_1225-0515 & 12 28 12.42 & -05 31 40.63 & \nodata & 15.584 & 14.947 & 15.524 & 15.767 & 15.118 & 0.005 & 2410 & 6210 & \nodata 1.96 & 0.52\\
HE\_1330-0354 & 13 33 10.67 & -04 10 05.80 & 15.248 & 15.000 & 14.380 & 14.937 & 15.169 & 14.537 & 0.082 & 1716 & 6257 & \nodata 2.29 & 1.05\\
HE\_2250-2132 & 22 53 40.48 & -21 16 23.96 & 14.905 & 14.392 & 13.662 & 14.220 & 14.526 & 13.735 & 0.027 & 1780 & 5705 & \nodata 2.22 & 0.41\\
HE\_2347-1254 & 23 50 10.01 & -12 37 50.46 & 13.839 & 13.358 & 12.779 & 13.248 & 13.499 & 12.833 & 0.064 & 783 & 6132 & \nodata 1.83 & 0.27\\
HE\_2347-1448 & 23 49 58.34 & -14 32 15.60 & 15.692 & 15.226 & 14.615 & 15.109 & 15.374 & 14.672 & 0.050 & 3062 & 6162 & \nodata 2.31 & 0.50\\
HE\_0244-4111 & 02 45 57.45 & -40 59 06.81 & 15.531 & 15.025 & 14.302 & 14.839 & 15.141 & 14.356 & 0.075 & 2380 & 5624 & \nodata 2.56 & 0.25\\
HE\_0441-4343 & 04 43 20.43 & -43 38 20.51 & \nodata & 15.559 & 14.835 & 15.388 & 15.695 & 14.904 & 0.047 & 3511 & 5629 & \nodata 2.52 & 0.33\\
HE\_0513-4557 & 05 15 12.21 & -45 54 10.46 & 16.279 & 15.743 & 15.045 & 15.609 & 15.913 & 15.133 & 0.038 & 3602 & 5629 & \nodata 2.79 & 0.39\\
HE\_0926-0508 & 09 28 55.35 & -05 21 40.48 & 12.340 & 12.194 & 11.617 & 12.126 & 12.357 & 11.735 & 0.086 & 457 & 6249 & \nodata 2.78 & 0.62\\
HE\_1006-2218 & 10 09 00.69 & -22 33 30.00 & \nodata & 13.773 & 13.197 & 13.717 & 13.936 & 13.345 & 0.102 & 928 & 6638 & \nodata 2.69 & 9.99\\
HE\_1015-0027 & 10 17 35.70 & -00 42 24.30 & 15.621 & 15.342 & 14.701 & 15.271 & 15.519 & 14.846 & 0.145 & 1886 & 6315 & \nodata 2.66 & 9.99\\
HE\_1120-0153 & 11 22 43.39 & -02 09 36.69 & 12.210 & 11.789 & 11.036 & 11.646 & 11.884 & 11.242 & 0.110 & 509 & 6191 & \nodata 2.77 & 0.63\\
HE\_1126-1735 & 11 28 51.39 & -17 51 42.82 & \nodata & 15.965 & 15.232 & 15.890 & 16.166 & 15.445 & 0.132 & 4118 & 5689 & \nodata 2.69 & 0.23\\
HE\_1413-1954 & 14 16 04.71 & -20 08 54.09 & \nodata & 15.235 & 14.593 & 15.171 & 15.418 & 14.745 & 0.225 & 1966 & 6533 & \nodata 3.22 & 1.45\\
HE\_2222-4156 & 22 25 28.65 & -41 40 57.72 & \nodata & 15.332 & 14.583 & 15.252 & 15.538 & 14.783 & 0.042 & 2773 & 5537 & \nodata 2.73 & 0.42\\
HE\_2325-0755 & 23 27 59.61 & -07 39 13.49 & 14.481 & 13.940 & 13.251 & 14.200 & 14.452 & 13.776 & 0.044 & 1600 & 5665 & \nodata 2.85 & 0.21\\
\enddata
\end{deluxetable*}
\end{rotatetable}

\section{Isochrones}\label{Isochrones}
To derive age we used two different sets of isochrones for old metal-poor populations: Padova isochrones\footnote{\url{http://stev.oapd.inaf.it/cmd}} and a Bag of Stellar Tracks and Isochrones\footnote{\url{http://basti-iac.oa-abruzzo.inaf.it/isocs.html}} (BaSTI).

We used these two sources of isochrones to check for systematics due to differences in stellar evolution models. For example, Padova isochrones do not take into account $\alpha$-enhancement while most of the studied stars are $\alpha$-enhanced. BaSTI isochrones instead have main sequence bluer than references globular clusters (\cite{Hidalgo2018}). We can notice that for lower metallicity ($\left[\frac{Fe}{H}\right] = -2.2$) the two sets of isochrones are in a good agreement in the turn-off point region, but BaSTI isochrones have slightly bluer in the main sequences and red giant branch is more vertical. Instead, for higher metallicity ($\left[\frac{Fe}{H}\right] = -1.2$) the shift between isochrones is more evident. BaSTI isochrones are fainter and redder. The reason for this shift could be that they take into account $\alpha$-enhancement that displaces isochrones towards higher metallicity (\cite{Salaris_1993}). Tab.\ref{tab:Parameters of isochrones} gives a summary of the isochrones we used in this study.

\begin{table}[]
    \centering
    \begin{tabular}{lll}
    \hline
    \hline
        Parameter& BaSTI & Padova\\
        \hline
        Age & 1 -- 15 Gyr with step 0.1 Gyr & 1 -- 20 Gyr with step 0.1 Gyr\\
        $\rm{[Fe/H]}$ & \nodata 1.05, -1.2, -1.3, -1.4, -1.55, -1.7, -1.9, -2.2, -2.5, -3.2 & \nodata 1.0 -- -2.2 dex with step 0.1 dex\\
        Heavy element mixture& $\rm{[\alpha/Fe]}$ = +0.4 &  = $\rm{[\alpha/Fe]}$=+0.0\\
        Photometric system & UBVIJHK, Gaia DR3 & UBVIJHK, Gaia DR3\\
        \hline
    \end{tabular}
    \caption{Characteristics of the BaSTI and Padova isochrones we used in this study.}
    \label{tab:Parameters of isochrones}
\end{table}

In order to use Padova isochrones properly, we need to correct them for $\alpha$-enhancement. To do so we used the technique proposed by \cite{Salaris_1993} where $\alpha$-enhanced isochrones can be reproduced by standard isochrones if a metallicity given by the following equation is used: 

\begin{equation}\label{alpha correction}
    Z=Z_0(0.638f_{\alpha}+0.362)
\end{equation}
where $f_{\alpha}=\frac{Z}{Z_{Sun}}=10^{\left[\frac{\text{element abundance}}{Fe}\right]}$

The result of $\alpha$-correction is shown in Fig.\ref{fig:Data distribution}. In this plot we can see that the correction for $\alpha$-enhancement increases the total metallicity on average by 0.2 dex. Moreover, Fig.\ref{fig:Metallicity all distances} indicates a good agreement between the stars' distribution and the isochrones, both colour-coded with metallicity.

\section{Reddening}\label{Reddening}
In this work we used two photometric systems: Gaia ($G$, $G_{BP}$, $G_{RP}$) and Johnson (B, V). Both of them cover the optical part of the spectrum thus they are quite sensitive to reddening. Therefore, correction for reddening is a critical step in our analysis. Our targets are located in the Galactic Halo ($|b| > 20^o$). Galactic Halo is poor in gas and dust and, as a consequence, it does not exhibit significant extinction. But even small reddening corrections might cause big uncertainties in age determination, that is why we used the 5 different sources of extinction values listed in Tab.\ref{tab:extinction sources} to get the best estimate for that effect.

\begin{table}[]
    \centering
    \begin{tabular}{lcccc}
        \hline
        \hline
        Source & Year & Type of the map & Coverage & Accuracy\\
        \hline
        \citeauthor{Schlegel_1998} & \citeyear{Schlegel_1998} & 2D & all sky & 16\%\\
        \citeauthor{Schlafly_2011} & \citeyear{Schlafly_2011} & 2D & all sky & 50 mmag\\
        \citeauthor{Queiroz_2019} (\texttt{StarHorse}) & \citeyear{Queiroz_2019}& 3D & $|Z_{Gal}| < 1$ kpc,
$R_{Gal} \lesssim 20$ kpc & 50-200 mmag\\
        \citeauthor{Green_2018} & \citeyear{Green_2018} & 3D & $\delta \gtrsim -30^o$, $d^\mathrm{*} \lesssim 60$ kpc & 10-100 mmag\\
        \citeauthor{Lallement_2018} & \citeyear{Lallement_2018} & 3D & $D^\mathrm{**} \leq 4000$ pc, $|Z_{Gal}| \leq 600$ pc & 10-150 mmag\\
        \citeauthor{Montalto_2021} & \citeyear{Montalto_2021} & method & $d \lesssim 2.5$ kpc & \\
        \hline
    \end{tabular}
    \caption{Extinction sources}
    \label{tab:extinction sources}
    \begin{footnotesize}
    \begin{itemize}
        \item[$^\mathrm{*}$] Distance from the Sun.
        \item[$^\mathrm{**}$] Distance from the Sun in the Galactic disk.
    \end{itemize}
    \end{footnotesize}

\end{table}

\cite{Schlegel_1998} is a full-sky 100 $\mu$m 2D map that is a reprocessed composite of the COBE/DIRBE and IRAS/ISSA maps, where the zodiacal foreground and confirmed point sources are removed. The uncertainty of the map is around 16\%.

\cite{Schlafly_2011} presented a full-sky 2D dust reddening map measured as the difference between the measured and predicted colours of a star, as derived from stellar parameters from the Sloan Extension for Galactic Understanding and Exploration Stellar Parameter Pipeline. They achieve uncertainties of 56, 34, 25, and 29 mmag in the colors u - g, g - r, r - i, and i - z, per star, though the uncertainty varies depending on the stellar type and the magnitude of the star.

\texttt{StarHorse} extinction values are computed by Bayesian isochrone fitting technique with stellar parameters ($T_{eff}$, log g, $\left[\frac{M}{H}\right]$) from spectroscopy, photometric magnitude ($m_{\lambda}$), parallax from Gaia DR2, and PARSEC isochrones. The extinction uncertainties are $\sim$70 mmag, when all photometric information is available, and $\sim$170 mmag if optical photometry is missing. For our data set the most common value of uncertainty is $\sim$200 mmag. (See also Sec. \ref{StarHorse distances}) They provide an coverage of the disc close to the Galactic mid-plane ($|Z_{Gal}| < 1 kpc$) from the Galactic center out to $R_{Gal} \sim 20 kpc$.

\cite{Green_2018} produced a new 3D map of interstellar dust reddening, covering three-quarters of the sky (declinations of $\delta \gtrsim -30^{o}$) out to a distance of several kiloparsecs. The map is based on high-quality stellar photometry of 800 million stars from Pan-STARRS 1 and 2MASS. They divide the sky into sightlines containing a few hundred stars each, and then infer stellar distances and types, along with the line-of-sight dust distribution. For our data set the mean uncertainty of this map is about 20 mmag.

\cite{Lallement_2018} selected low-reddening SDSS/APOGEE-DR14 red giants to obtain an empirical effective temperature- and metallicity-dependent photometric calibration in the Gaia G and 2MASS Ks bands. This calibration has been combined with Gaia G-band empirical extinction coefficients recently published, G, J, and Ks photometry, and APOGEE atmospheric parameters to derive the extinction of a large fraction of the survey targets. Distances were estimated independently using isochrones and the magnitude-independent extinction $K_{J-Ks}$. This new data set has been merged with the one used for the earlier version of the dust map. A new Bayesian inversion of distance-extinction pairs has been performed to produce an updated 3D map that covers $4000 \times 4000 \times 600$ $pc^3$ around the Sun. Uncertainty of this map for our data set varies from 10 to 150 mmag.

To extent \cite{Lallement_2018} map to farther distances we applied the method illustrated in \cite{Montalto_2021}. They suggested to use the dust distribution model of the Milky Way to calculate the amount of dust between the edge of the \cite{Lallement_2018} map and the real position of the object. It can be applied if in there is no specific dust and gas structures which is almost true for the Halo where all stars under consideration are located. In \cite{Montalto_2021} the extension method was applied for stars not farther than 2.5 kpc and tested to show good agreement with the photometric parameters.

\begin{figure}
\centering
\includegraphics[scale=0.5]{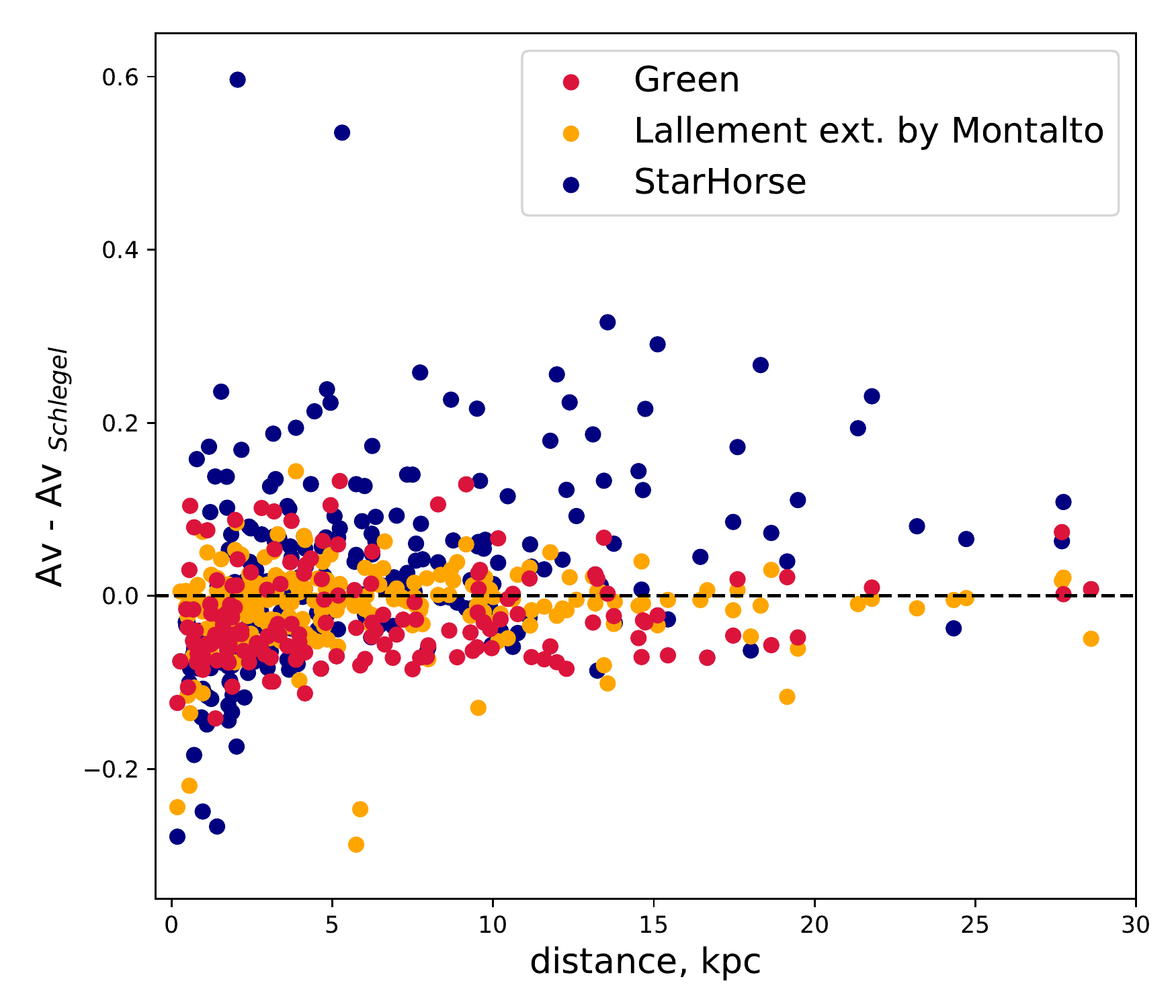}
\caption{Comparison of reddening from \cite{Green_2018}, \cite{Lallement_2018} extended by \cite{Montalto_2021} and \texttt{StarHorse} with reddening from \cite{Schlegel_1998} as a function of distance.}
\label{fig:Reddening comparison with Schlegel}
\end{figure}

\begin{figure}
\centering
\includegraphics[scale=0.5]{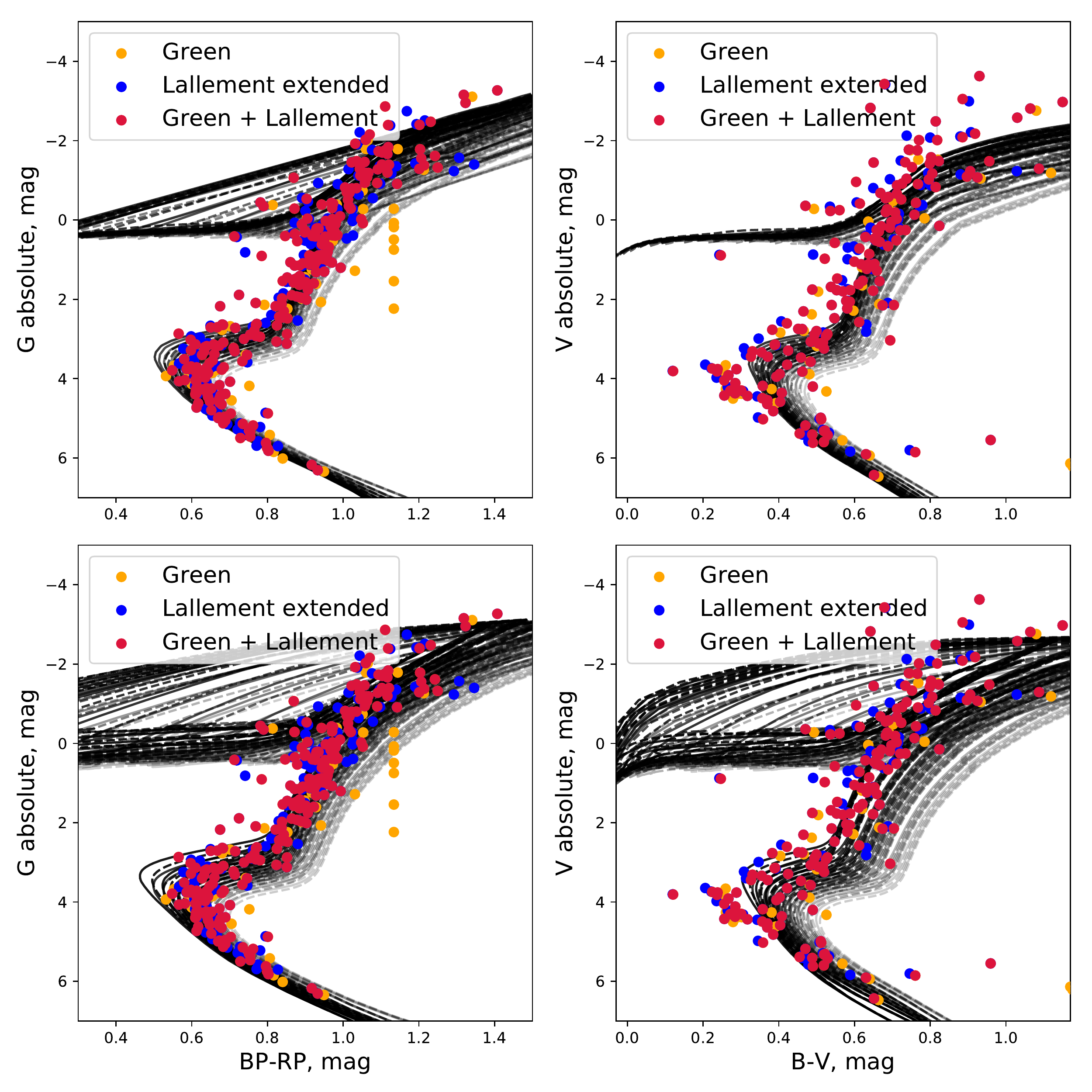}
\caption{CMD in Gaia (\textit{1$^{st}$ column}) and Johnson (\textit{2$^{nd}$ column}) photometry corrected using \cite{Lallement_2018} extended by \cite{Montalto_2021}, \cite{Green_2018} and the best correction value from one of the two (Green+Lallement). \textit{1$^{st}$ row}: Padova isochrones ([Fe/H]: -1.0 -- -2.2 dex, age: 10 -- 15 Gyr). \textit{2$^{nd}$ row}: BaSTI isochrones ([Fe/H]: -1.0 -- -3.2 dex, age: 10 -- 15 Gyr).}
\label{fig:Reddening GL}
\end{figure}

To get the best extinction estimate for our stars we studied all sources of reddening corrections listed in Tab.\ref{tab:extinction sources}. Since the range in distances for our stars is from 0.2 to 30 kpc (see Sec. \ref{Distance determination}) it is more precise to use a 3D map instead of 2D, which additionally has lower accuracy. In Fig.\ref{fig:Reddening comparison with Schlegel} we can see that \cite{Lallement_2018} extended by \cite{Montalto_2021} and \cite{Green_2018} are in a good agreement with each other and with \cite{Schlegel_1998} 2D map for far distances, for closer to the Sun star \cite{Schlegel_1998} map reddening is greater. \texttt{StarHorse} reddening instead shows a larger dispersion in all distance range and on average larger reddening values compared to \cite{Schlegel_1998} map. Therefore by comparing the three latest 3D maps we made a choice to use only \cite{Lallement_2018} extended by \cite{Montalto_2021} and \cite{Green_2018}. Both these maps have weak points. For example, \cite{Green_2018} map covers only three-quarters of the sky. \cite{Lallement_2018} does not have high accuracy in all directions. That is why our solution was to combine two reddening maps together by choosing for each star the best extinction estimate from \cite{Lallement_2018} extended by \cite{Montalto_2021} 3D map or from \cite{Green_2018} 3D map. The resulting reddening correction is shown in Fig.\ref{fig:Reddening GL}. We can see that the combined map shows the smallest stars' dispersion in the color-magnitude diagram (CMD). 

\section{Distance determination}\label{Distance determination}
Stellar distances constitute a fundamental quantity in astrophysics. In fact we need distances to compute absolute magnitudes for each metal-poor star in our data set. Distance is one of the crucial parameters that affect the obtained results. For this purpose we used four different techniques: Gaia DR3 parallaxes (\cite{GaiaDR3_2022}), Gaia EDR3 (\cite{GaiaEDR32021}) corrected by \cite{Lindegren2021}, the distances derived by \cite{Bailer-Jones2021}, and \cite{Queiroz_2019} (\texttt{StarHorse}).

\subsection{Gaia Data Release 3 (Gaia DR3) parallaxes}
The first main technique is parallax distance. Today we have very accurate trigonometric parallaxes (see for uncertainties Tab.\ref{tab:Unc GEDR3 astrometry}) obtained by the Gaia satellite for about 1,47 billion stars (\cite{GaiaDR3_2022}). From these trigonometric parallaxes, distance can be obtained through the following equation:
\begin{equation}\label{parallax distance}
    d = \frac{1}{\pi}
\end{equation}
where $d$ is the distance to the object in pc and $\pi$ is its trigonometric parallax in arcsecs. The problem with this distance determination is that, due to the structure of this equation, uncertainties on the distance are not symmetric around the mean value, especially for a large uncertainty on the parallax. That is why errors for each source must be computed separately for the higher and lower edges of the distance:

\begin{equation}\label{parallax distance error low}
    d_{low} = \frac{1}{\pi+\Delta\pi}
\end{equation}
\begin{equation}\label{parallax distance error high}
    d_{high} = \frac{1}{\pi-\Delta\pi}
\end{equation}

Uncertainties of Gaia Data Release 3 astrometry as given by the GAIA Collaboration are shown in Tab.\ref{tab:Unc GEDR3 astrometry}

\begin{table}[]
    \centering
    \begin{tabular}{lcccc}
        \hline
        \hline
        \multicolumn{1}{l}{Data product}&\multicolumn{4}{c}{Typical uncertainty}\\
        &G $<$ 15&G = 17 &G = 20 &G = 21 \\
        \hline
        Five-parameter astrometry&&&&\\
        \hline
        position, mas&0.01 - 0.02&0.05&0.4&1\\
        parallax, mas&0.02 -
        0.03&0.07&0.5&1.3\\
        \hline
        Six-parameter astrometry&&&&\\
        \hline
        position, mas&0.02 - 0.03&0.08&0.4&1\\
        parallax, mas&0.02 -     0.04&0.1&0.5&1.4\\
        \hline
    \end{tabular}
    \caption{Uncertainties of Gaia Data Release 3 astrometry from \cite{GaiaEDR32021}}
    \label{tab:Unc GEDR3 astrometry}
\end{table}

\subsection{Corrections for Gaia EDR3 parallaxes: \cite{Lindegren2021}}\label{Correction}

Parallaxes measured by \cite{GaiaEDR32021} can have some biases that have been measured by \cite{Lindegren2021}. \cite{Lindegren2021} found that parallaxes that correspond to quasars (distant objects, whose parallaxes should be distributed around zero) have a systematical offset from the expected distribution around zero, by a few tens of micro-arcsec. Based on quasars bias for faint sources they extended the map of the correction to lower magnitudes using physical pairs (binaries) and Large Magellanic Cloud sources. The parallax bias is found to depend in a non-trivial way on (at least) the magnitude, color, and ecliptic latitude of the source. Different dependencies apply to the five- and six-parameter solutions in Gaia EDR3. While it is not possible to derive a definitive recipe for the parallax correction, they give tentative expressions to be used at the researcher’s discretion and point out some possible paths toward future improvements. We applied the \cite{Lindegren2021} correction for downloaded Gaia EDR3 parallaxes and then computed distance and its lower and upper limit through Eq.(\ref{parallax distance}), (\ref{parallax distance error high}), (\ref{parallax distance error low}). The results are shown in Fig.\ref{fig:Metallicity all distances} second row.

\subsection{Corrections for Gaia EDR3 parallaxes: \cite{Bailer-Jones2021}}
Despite Gaia EDR3's high precision, the majority of stars observed by Gaia are distant or faint so their parallax uncertainties are large and this prevents the direct inversion of parallax for obtaining distance. That is why \cite{Bailer-Jones2021} used a probabilistic approach to estimate stellar distances that use a prior construction from a three-dimensional model of our Galaxy. This model includes interstellar extinction and Gaia's variable magnitude limit. They obtain two types of distances. The first, geometric, uses the parallax together with a direction-dependent prior on distance. The second, photo-geometric, additionally uses the color and apparent magnitude of a star, by exploiting the fact that stars of a given color have a restricted range of probable absolute magnitudes (plus extinction). Tests on simulated data and external validations show that the photo-geometric estimates generally have higher accuracy and precision for stars with poor parallaxes.

\subsection{\texttt{StarHorse} distances} \label{StarHorse distances}

An additional source of distances is the APOGEE DR16 \texttt{StarHorse} catalog where \cite{Queiroz_2019} combined spectroscopic (APOGEE-2 survey Data Release 16) and photometric (IR: 2MASS, AllWISE; Optical: PanSTARRS-1) data as well as parallaxes (Gaia Data Release 2). They used Bayesian isochrone-fitting code \texttt{StarHorse} to obtain distances and extinction for 388 815 APOGEE stars. All studied in this work stars are included. The typical distance uncertainties are $\sim6\%$ for APOGEE giants and $\sim2\%$ for APOGEE dwarfs. \texttt{StarHorse} uncertainties vary with the input spectroscopic catalogue, available photometry, and parallax uncertainties. Data are available at \textbf{\url{https://data.aip.de/projects/aqueiroz2020.html}}.

\subsection{Distance choice}\label{Distance choice}

\begin{figure}
\centering
\includegraphics[scale=0.4]{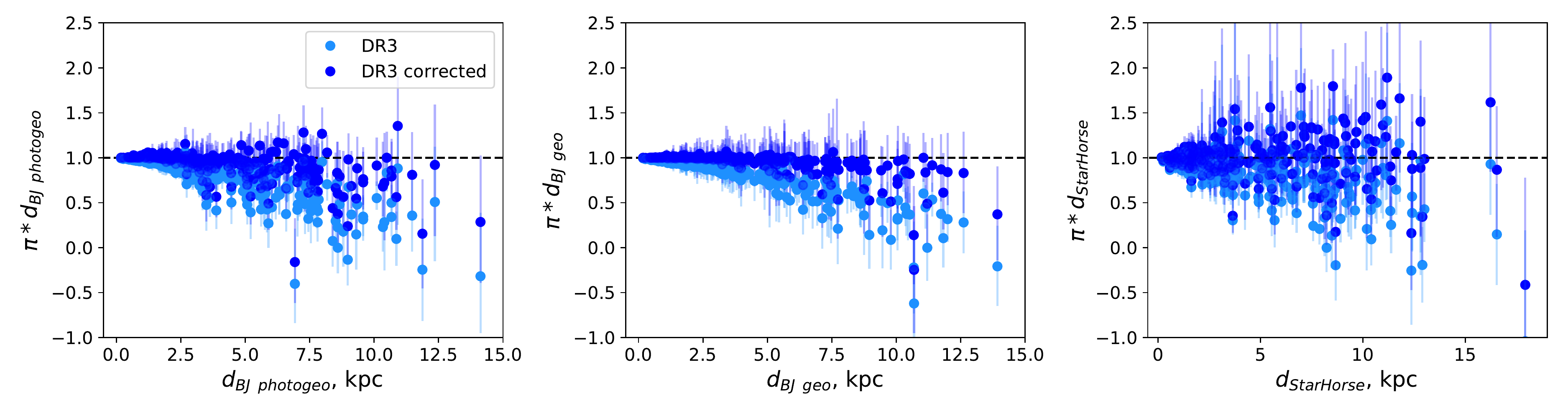}
\caption{Comparison of parallax from Gaia DR3 (light blue) and corrected Gaia EDR3 by \cite{Lindegren2021} (blue) with geometric (\textit{left}) and photogeometric (\textit{middle}) distance from \cite{Bailer-Jones2021} and with \texttt{StarHorse} distance from \cite{Queiroz_2019} (\textit{right}). The vertical axis shows the Gaia DR3 parallax and corrected parallax multiplied by the geometric distance: values under 1 correspond to the parallax distance being larger than the value of \cite{Bailer-Jones2021}/\texttt{StarHorse} distance and vice versa. The vertical error bars take into account the statistical uncertainties both on the parallax and the distance, but the horizontal error bars for the distance are not displayed.}
\label{fig:Comparison between distance and parallaxes}
\end{figure}

The comparison of distances from \cite{Bailer-Jones2021} and parallaxes from Gaia DR3 and corrected by \cite{Lindegren2021} are shown in Fig.\ref{fig:Comparison between distance and parallaxes} (left, middle panels). The vertical axis shows the Gaia DR3 parallax and corrected parallax multiplied by the geometric (Fig.\ref{fig:Comparison between distance and parallaxes}, left panel) and  photo-geometric (Fig.\ref{fig:Comparison between distance and parallaxes}, middle panel) distance: values under 1 correspond to the parallax distance larger than the value of \cite{Bailer-Jones2021} distance and vice versa. The vertical error bars take into account the statistical uncertainties both on the parallax and on the distance, but the horizontal error bars for the distance are not displayed. We can see that for close objects ($<$ 3 kpc) parallaxes corrected by \cite{Lindegren2021} are in good agreement with \cite{Bailer-Jones2021} geometric and photo-geometric distances. Beyond 3 kpc the corrected parallaxes give larger distances than the geometric and the photo-geometric ones. In general Gaia DR3 parallaxes yield larger distances in all ranges of distances. Some of the stars have negative parallaxes but positive distances can be derived from \cite{Bailer-Jones2021}.

\begin{figure}
\centering
\includegraphics[scale=0.3]{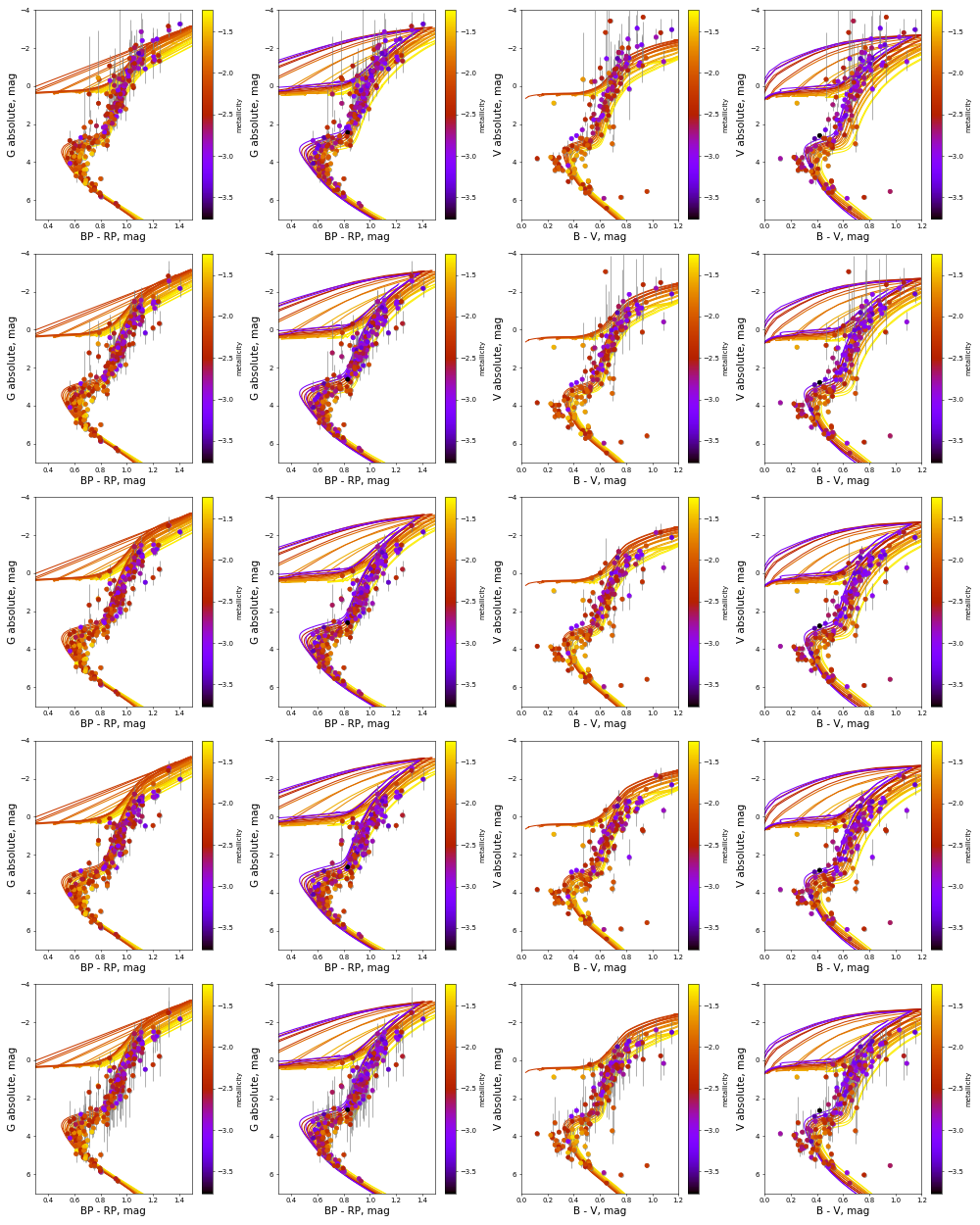}
\caption{Gaia (\textit{first two columns}) and Johnson (\textit{last two columns}) photometry color-coded with corrected metallicity for Padova isochrones (\textit{1$^{st}$, 3$^{rd}$ columns}) and with observed metallicity for BaSTI isochrones (\textit{2$^{nd}$, 4$^{th}$ columns}). Distances are computed by Gaia DR3 parallaxes (\textit{1$^{st}$ row}), Gaia EDR3 parallaxes corrected by \cite{Lindegren2021} (\textit{2$^{nd}$ row}), \cite{Bailer-Jones2021} (geo: \textit{3$^{rd}$ row}, photogeo: \textit{4$^{th}$ row}),  and computed by \cite{Queiroz_2019} (\texttt{StarHorse}) (\textit{5$^{th}$ row}). Isochrones are colour-coded with metallicity ([Fe/H]: -1.0 -- -2.2 dex, age: 10 -- 15 Gyr (Padova); [Fe/H]: -1.0 -- -3.2 dex, age: 10 -- 15 Gyr (BaSTI))}
\label{fig:Metallicity all distances}
\end{figure}

From Fig.\ref{fig:Comparison between distance and parallaxes} (right panel) we can notice that \texttt{StarHorse} distances and distances from parallax corrected by \cite{Lindegren2021} are in a good agreement. On the contrary, direct distances from Gaia EDR3 parallaxes show larger values. However, to consider the best estimate of distances for our data set a deeper investigations should be done using a CMD and isochrone fitting technique.
In Fig.\ref{fig:Metallicity all distances} stars are corrected for reddening with combined \cite{Green_2018} + \cite{Lallement_2018} extended by \cite{Montalto_2021} map and color-coded with corrected metallicity for Padova isochrones, observed - for BaSTI isochrones. Metallicity range for Padova isochrones: -1.3 -- -2.2 dex, for BaSTI isochrones: -1.3 -- -3.2 dex. For each metallicity 10 and 15 Gyr isochrone are plotted colour-coded with metallicity. Comparing different methods for distance determination we can highlight the following discrepancies, depending on the distance technique we use:

\begin{itemize}
    \item [-] Some stars are located in low probability regions, like above the MS and below the SGB, or on the red sie of the RGB. (Gaia DR3 parallaxes, Lindegren, Bailer-Jones, \texttt{StarHorse})
    \item [-] Non negligible shift in the position of RGB and TO stars in comparison with the isochrones of the same metallicity. (Lindegren, Bailer-Jones, \texttt{StarHorse})
    \item [-] Different shape of the targets RGB if compared with the shape of isochrone RGB. (Lindegren, Bailer-Jones, \texttt{StarHorse})
\end{itemize}

According to the listed above criteria, we can conclude that Gaia parallaxes show the best fit to the data. In this case we can see also that fewer targets only are lying in the low probability region. Moreover, all stars with metallicities lower than the lower limit for isochrones metallicity range lie above RGB where the estimated position of these stars should be. Lindegren correction and \texttt{StarHorse} distances move low metallicity stars in RGB to the position of isochrones with higher metallicity and break the shape of RGB. Bailer-Jones additionally shift more stars to the low probability region which reduces the quality of the data set. All these tests show that the best distance estimate estimator for our purposes is the distance obtained directly inverting Gaia parallaxes.

A possible explanation for the lower accuracy we obtained applying \cite{Lindegren2021} correction is that they are based on faint sources and, as a consequence, they are less accurate for brighter magnitudes as it is the case for our targets. Bailer-Jones also used Lindegren correction as an input. Because of this, their corrections are affected by the same bias. \texttt{StarHorse} distances are less precise instead because \cite{Queiroz_2019} used Gaia Data Release 2 parallaxes as prior for the distance calculation, which have lower accuracy compared with new Gaia DR3 parallaxes.

\section{Derivation of the absolute magnitude}
To be able to compare isochrones with observational data we need to convert observed G photometry to its absolute magnitude. To do this we need to apply distance modulus and reddening correction using:

$$M_{\lambda} = m_{\lambda} + 5 - 5\cdot log(d) - A_{\lambda}$$

where $d$ is the distance to the star in parsecs and $A_{G} = A_{V}\cdot coef$ (for $coef$ see Tab.\ref{tab:reddening coefficients}). These transformations assume the standard reddening law $A_{V} = R_{V}\cdot E(B-V)$ with $R_{V}=3.1$ since all our targets are far from the galactic plane or the galactic Bulge, where $R_{V}$ can assume different values. 

\begin{table}[]
    \centering
    \begin{tabular}{lcccccc}
    \hline
    \hline
        &B&V&I&G&$G_{BP}$&$G_{RP}$\\
        \hline
        $A_{\lambda}/A_{V}$&1.326&1.000&0.599&0.861&1.061&0.648\\
        \hline
    \end{tabular}
    \caption{Reddening coefficients for different photometric filters}
    \label{tab:reddening coefficients}
\end{table}

\section{Difference in CMDs for different photometric systems}
A closer look at the data evidences additional peculiarities. Fig.\ref{fig:Metallicity all distances} shows that part of TO stars in the Johnson photometry (the two columns on the right) are blue-shifted with respect to the isochrones and also if compared with their position in the Gaia photometry (the two columns on the left). In order to investigate deeper this behaviour we performed further tests.

\begin{figure}
\centering
\includegraphics[scale=0.5]{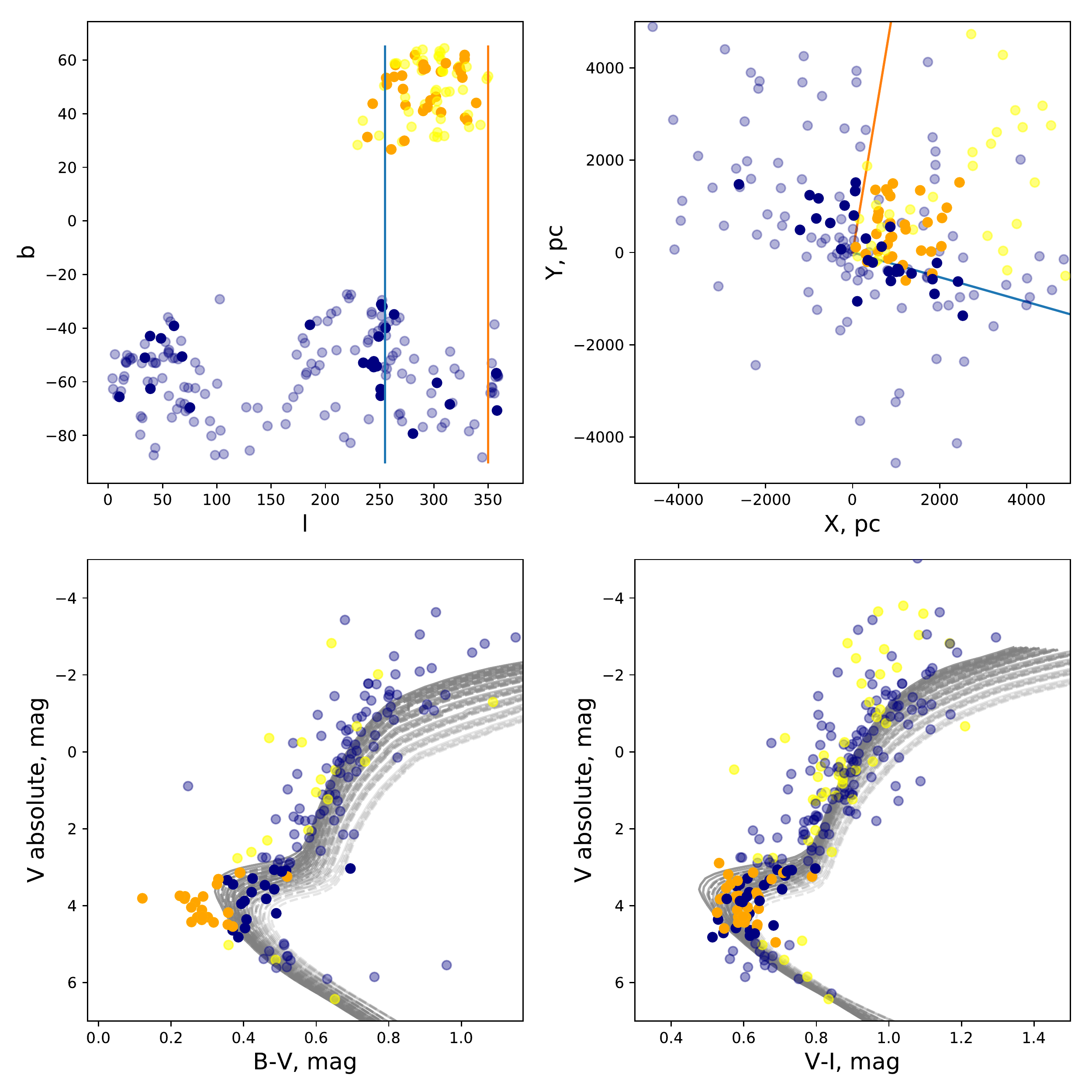}
\caption{\textit{Top}: Position of our targets on the sky (\textit{left}, Galactic coordinates) and in the Galactic plane (\textit{right}), color-coded with Galactic latitude ($b > 0$: yellow, $b < 0$: blue). \textit{Bottom}: CMD in V vs. B-V (\textit{left}) and V vs. V-I (\textit{right}) filters, yellow points are non-TO stars with $b>0$, orange points are TO stars with $b>0$, light blue are non-TO stars with $b<0$, while dark blue are TO stars with $b<0$. Padova isochrones are added as a reference ([Fe/H]: -1.0 -- -2.2 dex, age: 10 -- 15 Gyr).}
\label{fig:bl XY V_VI}
\end{figure}

First of all, we colour-coded our data with Galactic latitude (Fig.\ref{fig:bl XY V_VI}, top row). The bottom left panel of this figure report the position of the targets on the V vs. B-V CMD. It is clear that all the stars with the larger TO-shift are located at positive Galactic latitude, while all the other stars have negative Galactic latitudes. 
The difference in position might indicate a difference in reddening, but since our data set is corrected for it, only a particularly improbable kind of extinction could explain this behaviour.
On the other hand Fig.\ref{fig:bl XY V_VI} (bottom right panel) reports the V vs. V-I CMD, where we do not see such an affect. This means that the shift is entirely due to the B filter.

\subsubsection{Chemical analysis} \label{chemical analysis}
\begin{figure}
\centering
\includegraphics[scale=0.5]{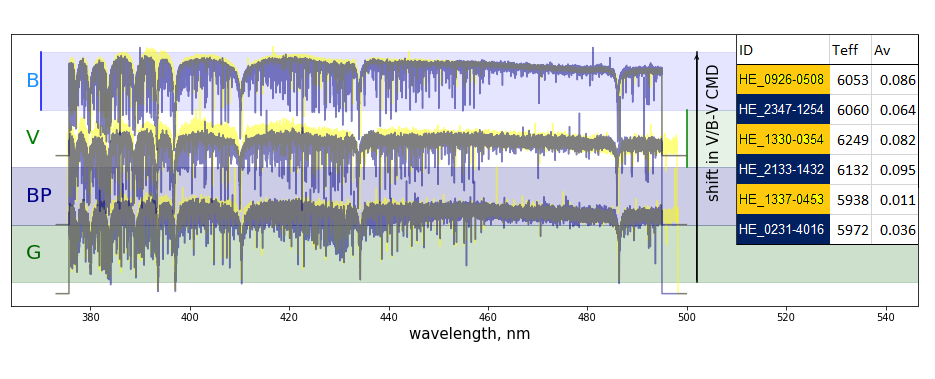}
\caption{Comparison of the spectra for bluer and redder stars in $V/B_V$ CMD. Gaia and Johnson photometric filters ranges are plotted with color.}
\label{fig:Spectra comparison}
\end{figure}

\begin{figure}
\centering
\includegraphics[scale=0.5]{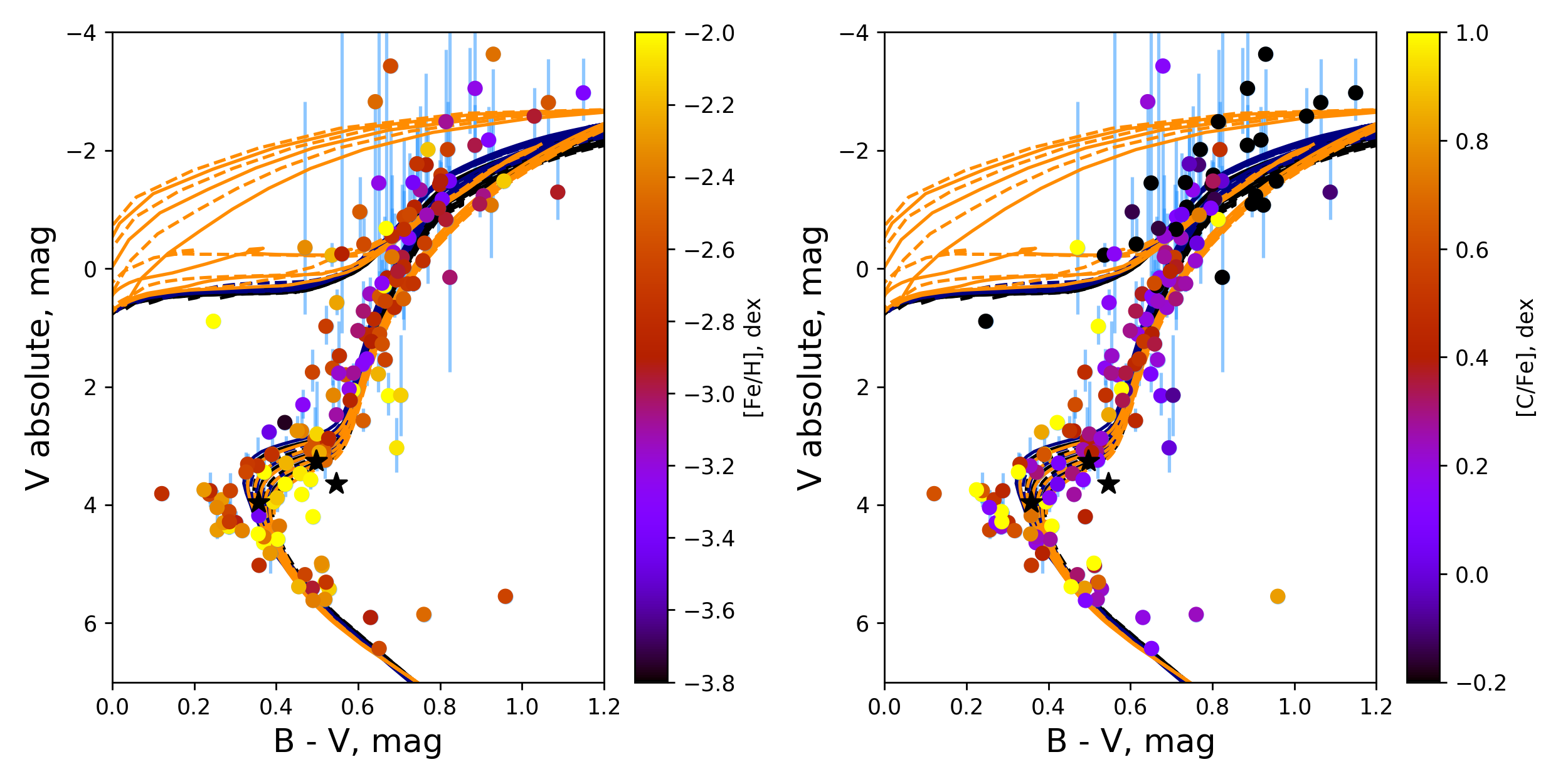}
\caption{Johnson photometry color-coded with $\rm[Fe/H]$ (\textit{left panel}) and $\rm[C/Fe]$ (\textit{right panel}). Distance computed by means of Gaia DR3 parallaxes. Black stars are stars: HD 84937, HD 132475, and HD 140283 (\cite{VandenBerg2014}). Padova (blue) and BaSTI (orange) isochrones with [Fe/H] = -2.2 dex, age: 10 -- 15 Gyr are added as a reference.}
\label{fig:FeC}
\end{figure}

One of the reasons why stars with the same temperature (as inferred from the BP-GP and V-I colors) and luminosity (as inferred from G and V magnitudes) differ in just one of the filters, is that this filter contains strong molecular bands and that the elements responsible for these band (usually C,N,O) vary their abundances from one star the other. Also chemical abundances variation of C, N, O, Ne, Mg, Si, S, Ca, Fe affect opacity of the star and change the continuum emission in that part of the spectrum (\cite{Salaris_1993}). To check these effects we download from the ESO database\footnote{\url{http://archive.eso.orG vs. scienceportal/home}} flux-calibrated UVES spectra for couples of stars with the same T$_{eff}$ and similar reddening ($\Delta A_{V} < 0.02$ mag) but different dereddened B-V colors.
In Fig.\ref{fig:Spectra comparison} we can see that when the distance between shifted (yellow) and non-shifted (blue) stars increases, also the continuum of the shifted stars increases. It happens specifically in the range of the B filter. To check the dependence between the shift and chemical composition, we colour-coded CMDs with abundances of C (the only element responsible of molecular bands we have) and Fe. From Fig.\ref{fig:FeC} we can see that shifted stars are enhanced in C and more metal-poor on average. Since we see no presence of strong molecular bands in the spectra, we conclude it is the change in the opacity of the atmospheres of the stars due to the different C and Fe contents that is the cause of shift in the B-V color. Specifically higher [C/Fe] abundance and lower [Fe/H] content increase the flux in the B filter.

For this reason Johnson photometry can not be properly used for age determination for very metal-poor stars. Gaia filters are wider and so less affected by this phenomenon and it is the only photometry we will use for the age determination.

\section{Age determination method}
To derive age we developed an automatic technique based on isochrone fitting. First, we filtered our data with the following criteria:
\begin{itemize}
    \item [1.] Distance $>0$
    \item [2.] Distance error $<20\%$
    \item [3.] Metallicity range: -2.3 –- -1 dex for Padova isochrone; -3.3 -- -1 dex for BaSTI isochrone
    \item [4.] Absolute magnitude is in TO-point region: $2.4<G<4.7$ mag
\end{itemize}

The first two criteria exclude stars with bad astrometric measurements. The third cutoff is based on the metallicity range covered by isochrone data sets. We should note that for different isochrones the cutoff is different. And the fourth cutoff leaves only stars from the TO-point region - the part of the CMD most sensitive to age.
The algorithm takes a set of isochrones matching the metallicity of the target and searches for the closest. Each isochrone is define by a set of X,Y points in the 2D space of the CMD. The distance used to choose the closest isochrone is the shortest between the perpendicular to one of the segments defines by two consecutive X,Y points, and the distance to the closest point of the isochrone. 
The age of the closest isochrone is then considered as a fisrt guess of age of the star if the distance is less than $10^{-4}$ mag (the maximum distance between isochrones with age step 100 Myr). Otherwise, age can not be derived, since the star is too far from all isochrones.

As a second step, we consider the mean parallax, reddening, metallicity, and photometry of each target and the related errors. Assuming a gaussian distribution we randomly distributed 10.000 points in the parameter space and applied the age determination technique to each of them.
The result is an age histogram that is fitted with a Gaussian function. We consider its mean as the age of the star and $\sigma$ as its uncertainty. 

\begin{figure}
\centering
\includegraphics[scale=0.6]{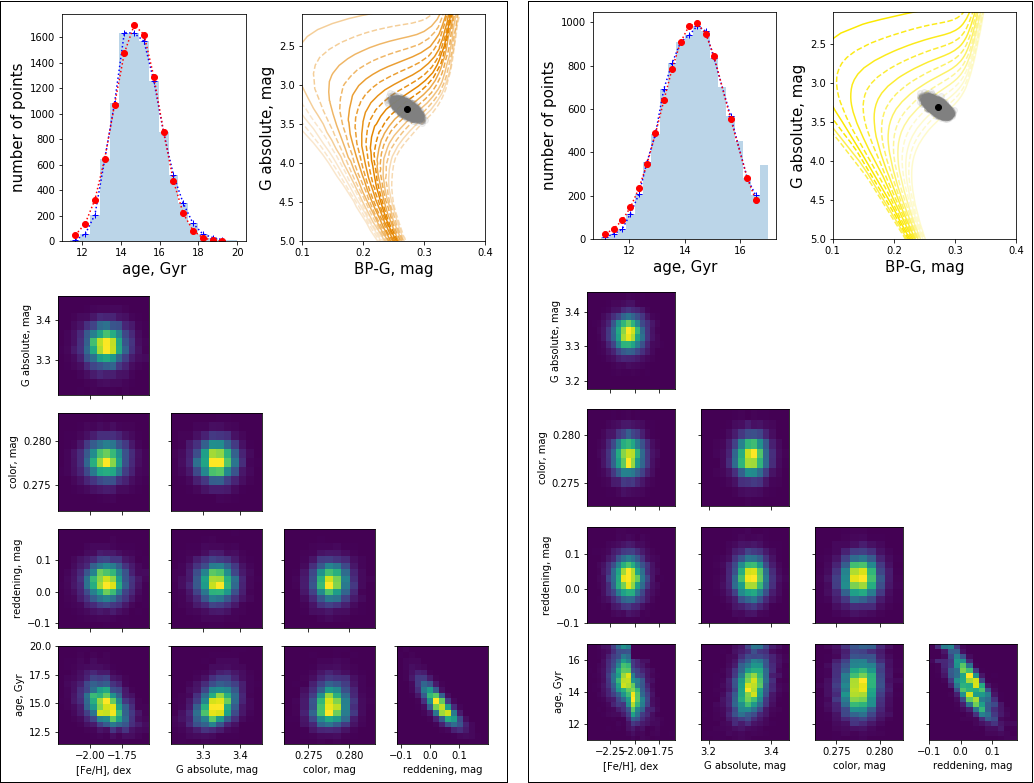}
\caption{Resulting plot of age determination from Padova (\textit{left panel}) and BaSTI (\textit{right panel}) isochrones for star HE 0023$-$4825 in diagrams G vs. BP-G. Grey dotes in the CMD are randomly selected input parameters. Black dote is original stellar parameters.}
\label{fig:res age}
\end{figure}

In Fig.\ref{fig:res age} we can see an illustration of the age determination for Padova and BaSTI isochrones. Isochrones have an upper limit of 20 Gyr for Padova, and from 15.4 to 19.2 Gyr (depending on metallicities) for BaSTI. Because of this, all stars redder than the oldest isochrone are considered to have the age of this isochrone. Therefore, we get some saturation in the bin corresponding to the oldest isochrone. We cut saturated bins as not representative. We derived age in three different combinations of photometric filters: G vs. G$_{BP}$-G$_{RP}$, G vs. G$_{BP}$-G, G vs. G-G$_{RP}$.

To get the best age sample we applied the following criteria:

\begin{itemize}
    \item [1.] Age derived in all three filter combinations: G vs. G$_{BP}$-G$_{RP}$, G vs. G$_{BP}$-G, G vs. G-G$_{RP}$.
    \item [2.] Derived age is not closer to the oldest isochrone less than 1$\sigma$.
\end{itemize}

In this way we reduced the targets for which we can derive good ages to 28.

\begin{longrotatetable}
\begin{deluxetable*}{lllllllllllllllll}
\tablecaption{Age of metal-poor stars.\label{tab:age}}
\tablewidth{700pt}
\tabletypesize{\scriptsize}
\tablehead{ID&$\left[\frac{Fe}{H}\right]$&\multicolumn{6}{c}{age from Padova isochrones, Gyr}&\multicolumn{6}{c}{age from BaSTI isochrones, Gyr}&\multicolumn{3}{c}{average, Gyr}\\
&&\multicolumn{2}{c}{G vs. G$_{BP}$-G}&\multicolumn{2}{c}{G vs. G-G$_{RP}$}&\multicolumn{2}{c}{G vs. G$_{BP}$-G$_{RP}$}&\multicolumn{2}{c}{G vs. G$_{BP}$-G}&\multicolumn{2}{c}{G vs. G-G$_{RP}$}&\multicolumn{2}{c}{G vs. G$_{BP}$-G$_{RP}$} & Padova & BaSTI & Padova+BaSTI\\
&dex&age&error&age&error&age&error&age&error&age&error&age&error&age&age&age}

\startdata
HE 0023-4825 & -2.06 & 14.8 & $\pm$1.2 & 14.9 & $\pm$1.1 & 14.9 & $\pm$1.1 & 14.4 & $\pm$1.2 & 14.4 & $\pm$1.2 & 14.4 & $\pm$1.2 & 14.9 & 14.4 & 14.6\\
HE 0109-3711 & -1.91 & 11.8 & $\pm$2.3 & 11.9 & $\pm$2.4 & 11.8 & $\pm$2.3 & 11.6 & $\pm$2.7 & 11.8 & $\pm$2.7 & 11.7 & $\pm$2.7 & 11.8 & 11.7 & 11.8\\
HE 0231-4016 & -2.08 & 13.3 & $\pm$1.3 & 13.3 & $\pm$1.2 & 13.3 & $\pm$1.3 & 15.4 & $\pm$1.3 & 15.4 & $\pm$1.2 & 15.4 & $\pm$1.2 & 13.3 & 15.4 & 14.4\\
HE 0340-3430 & -1.95 & 13.6 & $\pm$0.9 & 13.6 & $\pm$0.9 & 13.6 & $\pm$0.9 & 12.9 & $\pm$0.9 & 13.1 & $\pm$0.9 & 13.1 & $\pm$0.9 & 13.6 & 13.0 & 13.3\\
HE 0430-4404 & -2.07 & 7.7 & $\pm$3.0 & 6.6 & $\pm$2.7 & 7.0 & $\pm$2.8 & 13.8 & $\pm$2.8 & 12.7 & $\pm$2.5 & 13.2 & $\pm$2.6 & 7.1 & 13.2 & 10.1\\
HE 0447-4858 & -1.69 & 14.0 & $\pm$2.1 & 13.4 & $\pm$1.7 & 13.6 & $\pm$1.9 & 13.1 & $\pm$2.1 & 12.7 & $\pm$1.9 & 12.9 & $\pm$2.0 & 13.7 & 12.9 & 13.3\\
HE 0501-5139 & -2.38 & 9.0 & $\pm$2.3 & 9.0 & $\pm$2.3 & 9.0 & $\pm$2.3 & 8.9 & $\pm$2.4 & 8.9 & $\pm$2.4 & 8.9 & $\pm$2.4 & 9.0 & 8.9 & 8.9\\
HE 0519-5525 & -2.52 & 13.3 & $\pm$1.8 & 13.6 & $\pm$1.8 & 13.5 & $\pm$1.8 & 13.3 & $\pm$1.6 & 13.6 & $\pm$1.6 & 13.5 & $\pm$1.6 & 13.5 & 13.4 & 13.5\\
HE 0534-4615 & -2.01 & 16.0 & $\pm$2.9 & 16.0 & $\pm$2.9 & 16.0 & $\pm$2.9 & 13.8 & $\pm$2.3 & 13.8 & $\pm$2.2 & 13.9 & $\pm$2.2 & 16.0 & 13.8 & 14.9\\
HE 0938+0114 & -2.51 & 12.4 & $\pm$1.9 & 13.8 & $\pm$1.8 & 13.2 & $\pm$1.8 & 13.8 & $\pm$1.7 & 14.6 & $\pm$1.0 & 14.3 & $\pm$1.2 & 13.1 & 14.2 & 13.7\\
HE 1052-2548 & -2.29 & 15.3 & $\pm$0.7 & 16.1 & $\pm$0.6 & 15.8 & $\pm$0.6 & 15.2 & $\pm$0.5 & 16.0 & $\pm$0.4 & 15.7 & $\pm$0.4 & 15.7 & 15.7 & 15.7\\
HE 1105+0027 & -2.42 & 11.2 & $\pm$2.1 & 11.4 & $\pm$2.2 & 11.3 & $\pm$2.1 & 12.8 & $\pm$2.4 & 12.9 & $\pm$2.4 & 12.8 & $\pm$2.4 & 11.3 & 12.8 & 12.1\\
HE 1225-0515 & -1.96 & 14.6 & $\pm$1.6 & 14.6 & $\pm$1.6 & 14.6 & $\pm$1.6 & 14.7 & $\pm$1.8 & 14.6 & $\pm$1.6 & 14.7 & $\pm$1.8 & 14.6 & 14.7 & 14.6\\
HE 1330-0354 & -2.29 & 12.3 & $\pm$0.9 & 13.3 & $\pm$0.8 & 12.9 & $\pm$0.8 & 13.7 & $\pm$0.9 & 14.6 & $\pm$0.8 & 14.2 & $\pm$0.8 & 12.8 & 14.2 & 13.5\\
HE 2250-2132 & -2.22 & 13.2 & $\pm$1.2 & 13.6 & $\pm$1.3 & 13.4 & $\pm$1.3 & 12.8 & $\pm$1.1 & 13.2 & $\pm$1.3 & 13.0 & $\pm$1.2 & 13.4 & 13.0 & 13.2\\
HE 2347-1254 & -1.83 & 14.6 & $\pm$1.7 & 14.6 & $\pm$1.5 & 14.6 & $\pm$1.6 & 14.5 & $\pm$1.6 & 14.6 & $\pm$1.4 & 14.4 & $\pm$1.4 & 14.6 & 14.5 & 14.6\\
HE 2347-1448 & -2.31 & 8.7 & $\pm$1.8 & 8.8 & $\pm$1.8 & 8.8 & $\pm$1.8 & 8.4 & $\pm$1.8 & 8.5 & $\pm$1.9 & 8.5 & $\pm$1.9 & 8.8 & 8.5 & 8.6\\
HE 0244-4111 & -2.56 & \nodata & \nodata & \nodata & \nodata & \nodata & \nodata & 12.6 & $\pm$1.3 & 13.0 & $\pm$1.3 & 12.7 & $\pm$1.3 & \nodata & 12.8 & \nodata \\
HE 0441-4343 & -2.52 & \nodata & \nodata & \nodata & \nodata & \nodata & \nodata & 10.3 & $\pm$1.9 & 10.5 & $\pm$1.9 & 10.4 & $\pm$1.9 & \nodata & 10.4 & \nodata \\
HE 0513-4557 & -2.79 & \nodata & \nodata & \nodata & \nodata & \nodata & \nodata & 12.5 & $\pm$2.5 & 12.4 & $\pm$2.5 & 12.4 & $\pm$2.5 & \nodata & 12.4 & \nodata \\
HE 0926-0508 & -2.78 & \nodata & \nodata & \nodata & \nodata & \nodata & \nodata & 14.6 & $\pm$0.8 & 14.9 & $\pm$0.6 & 14.8 & $\pm$0.7 & \nodata & 14.8 & \nodata \\
HE 1006-2218 & -2.69 & \nodata & \nodata & \nodata & \nodata & \nodata & \nodata & 12.7 & $\pm$1.1 & 13.1 & $\pm$0.8 & 12.9 & $\pm$0.9 & \nodata & 12.9 & \nodata \\
HE 1015-0027 & -2.66 & \nodata & \nodata & \nodata & \nodata & \nodata & \nodata & 14.9 & $\pm$1.3 & 15.6 & $\pm$1.0 & 15.1 & $\pm$1.0 & \nodata & 15.2 & \nodata \\
HE 1120-0153 & -2.77 & \nodata & \nodata & \nodata & \nodata & \nodata & \nodata & 10.7 & $\pm$0.7 & 11.0 & $\pm$0.6 & 10.8 & $\pm$0.6 & \nodata & 10.8 & \nodata \\
HE 1126-1735 & -2.69 & \nodata & \nodata & \nodata & \nodata & \nodata & \nodata & 9.4 & $\pm$2.6 & 9.5 & $\pm$2.6 & 9.4 & $\pm$2.6 & \nodata & 9.4 & \nodata \\
HE 1413-1954 & -3.22 & \nodata & \nodata & \nodata & \nodata & \nodata & \nodata & 13.1 & $\pm$0.9 & 15.3 & $\pm$1.5 & 14.2 & $\pm$1.2 & \nodata & 14.2 & \nodata \\
HE 2222-4156 & -2.73 & \nodata & \nodata & \nodata & \nodata & \nodata & \nodata & 13.3 & $\pm$2.4 & 13.7 & $\pm$2.4 & 13.5 & $\pm$2.4 & \nodata & 13.5 & \nodata \\
HE 2325-0755 & -2.85 & \nodata & \nodata & \nodata & \nodata & \nodata & \nodata & 13.3 & $\pm$1.2 & 13.6 & $\pm$1.2 & 13.6 & $\pm$1.2 & \nodata & 13.5 & \nodata \\
\enddata
\end{deluxetable*}
\end{longrotatetable}

\section{Results}\label{results}

\subsection{Age}
\begin{figure}
\centering
\includegraphics[scale=0.5]{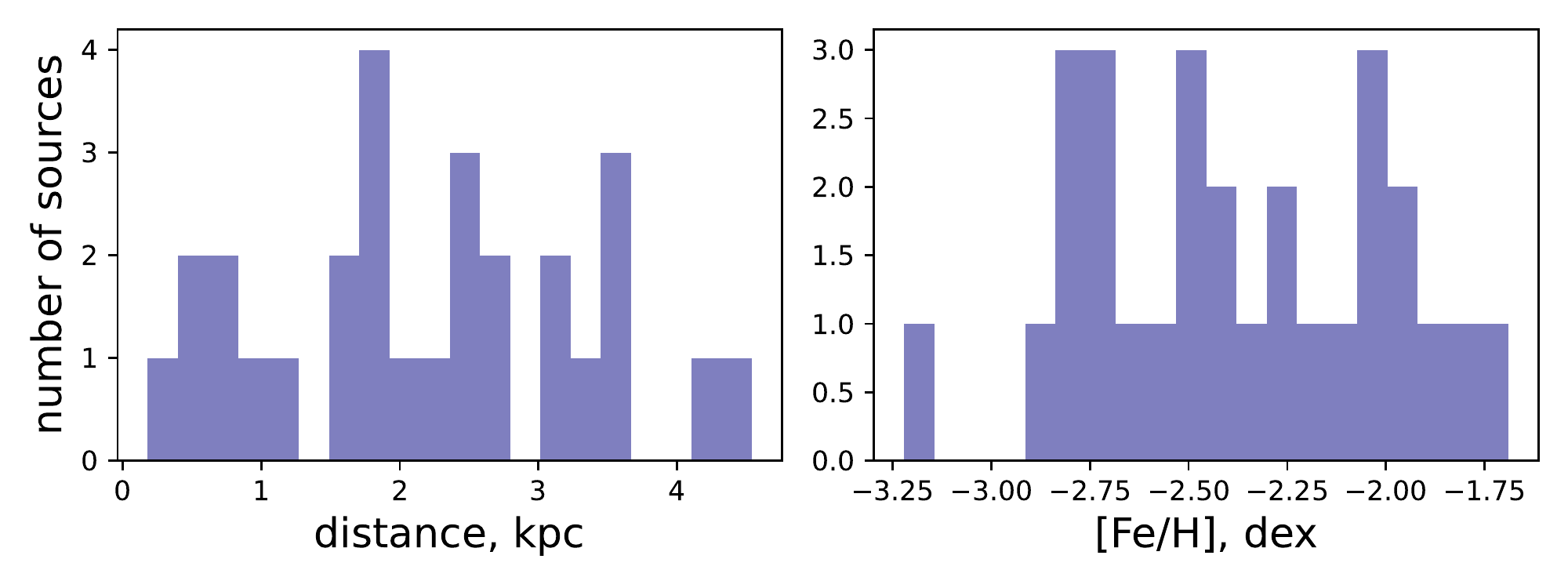}
\caption{Distribution of the distance and metallicity for the 28 stars with derived age.}
\label{fig:dist met 28}
\end{figure}

\begin{figure}
\centering
\includegraphics[scale=0.5]{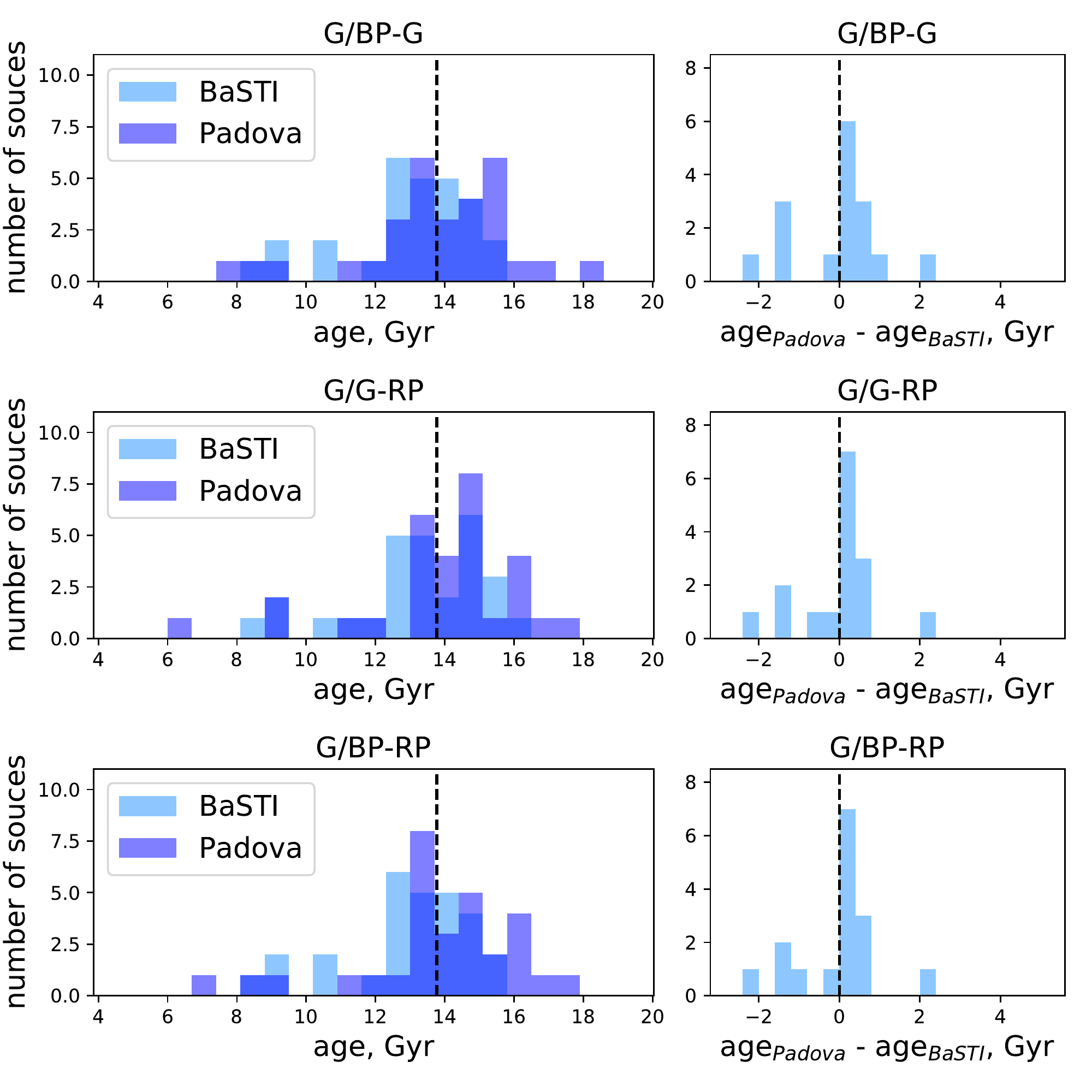}
\caption{\textit{Left}: dispersion in age determination from Padova and BaSTI isochrones in diagrams G vs. G$_{BP}$-G, G vs. G-G$_{RP}$, G vs. G$_{BP}$-G$_{RP}$. Vertical dotted line is the age of the Universe 13.77 Gyr (\cite{Bennett2013}). \textit{Right}: dispersion in age difference between Padova and BaSTI isochrones in diagrams G vs. G$_{BP}$-G, G vs. G-G$_{RP}$, G vs. G$_{BP}$-G$_{RP}$. Vertical dotted line is a zero point.}
\label{fig:age}
\end{figure}

The derived ages for 28 selected stars are reported in Tab.\ref{tab:age}. The distribution of the distance nd metallicity is reported in the Fig.\ref{fig:dist met 28}.Out of 28 we have only 17 stars for which age is derived from both isochrone sets and for 11 stars age was derived only from BaSTI isochrone set due to its lower limit in metallicity. In Fig.\ref{fig:age} on the left we can see the dispersion of the age, where the vertical dotted line show the age of the Universe 13.77 Gyr (\cite{Bennett2013}). On the right of Fig.\ref{fig:age} shows that on average Padova isochrones show older age but the systematic difference is much less than 1 Gyr. 

\begin{figure}
\centering
\includegraphics[scale=0.5]{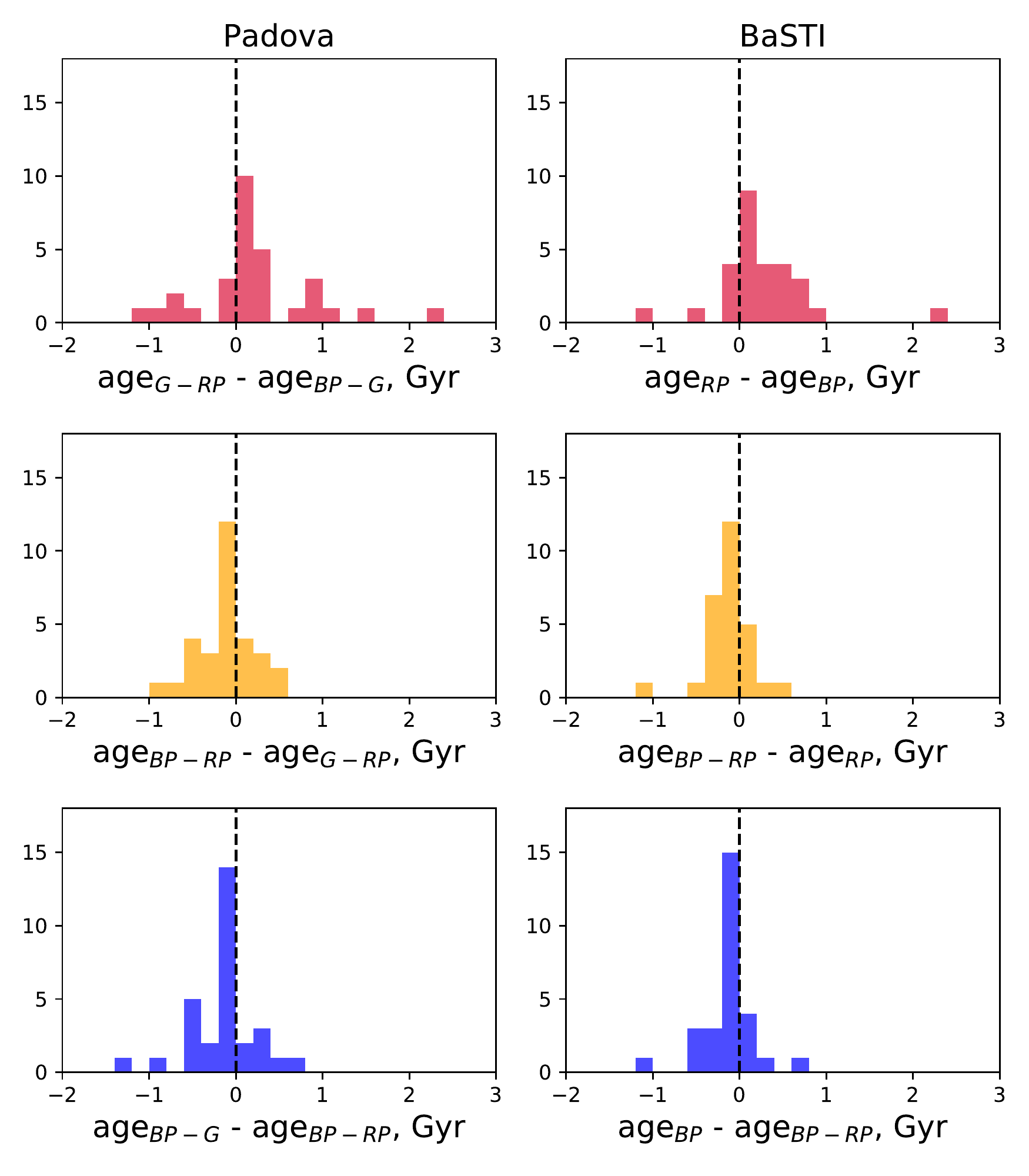}
\caption{Dispersion of difference in age from different filter combinations: G vs. G$_{BP}$-G, G vs. G-G$_{RP}$, G vs. G$_{BP}$-G$_{RP}$ for Padova and BaSTI isochrones. Vertical dotted line is zero-line.}
\label{fig:age diffilters}
\end{figure}

We also compared ages derived with different filter combinations: G vs. G$_{BP}$-G, G vs. G-G$_{RP}$, G vs. G$_{BP}$-G$_{RP}$. The result is reported in Fig.\ref{fig:age diffilters} where we can notice that the peak of all distributions is compatible with zero within the errors. This mean that ages derived with different filter combinations are the same.

\subsection{Comparison with old metal-poor globular clusters: M30, M92, and NGC 6397}\label{comparison with GCs}
We used metal-poor globular clusters (GCs) as a test of our age determination technique. We calculated the age for each star of GC located in the TO-point region ($2.4 < G < 4.7$ mag). All parameters for age determination of GC are listed in Tab.\ref{tab:GC}. Parallax was derived as a mean parallax from Gaia DR3 of all members of the cluster. All three GC are located far from the Sun ($> 2.5 kpc$). They lie outside the \cite{Green_2018} reddening map, and far from the edge of the \cite{Lallement_2018} reddening map that is why we choose extinction coefficients from 2D \cite{Schlegel_1998} map.  Metallicities are taken as an average of results listed in the SIMBAD catalog\footnote{\url{http://simbad.u-strasbg.fr/simbad/sim-fbasic}}. Reference ages were collected from \citet{Correnti_2018} (NGC 6397), \citet{Kains_2013} (M 30), \citet{VandenBerg_2016} (M 92).

\begin{figure}
\centering
\includegraphics[scale=0.5]{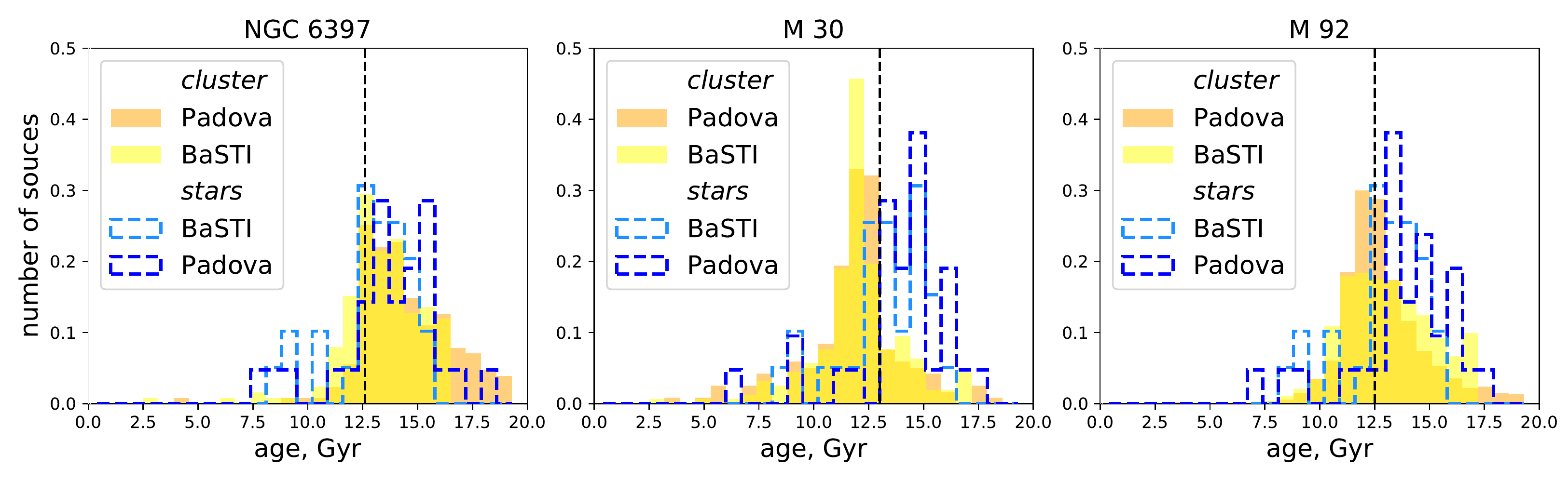}
\caption{Dispersion of the age for GC NGC 6397, M 30, M92 compared with metal-poor stars under investigtion.}
\label{fig:GC}
\end{figure}

Firstly, we checked the accuracy of our automatic age determination method by comparing the age derived by different authors with our results. As we can notice in the Fig.\ref{fig:GC} that the age determined by our method is in a good agreement with ages derived by other authors. Only for M 30 age is slightly younger. This inconsistency is caused by uncertainties of input data, especially in parallax and reddening.

Secondly, we compare the distribution of the age of the GC with the age distribution of metal-poor stars. In Fig.\ref{fig:GC} we can see that on average our metal-poor stars are older than all three GCs by about 1 Gyr. Also, it is worth mentioning that the width of the GC's age distribution shows that stars whose age is older than the age of the universe can be explained within the natural dispersion of the parameters in the same way as in GC.

\begin{table}[]
    \centering
    \begin{tabular}{lcccccc}
    \hline
    \hline
        Name&$\rm[Fe/H]$&age&$\pi$&d&A$_v$&age(ref.)\\
        &dex&Gyr&mas&pc&mag&Gyr\\
        \hline
        NGC 6397& -1.99& 12.9 &0.397& 2519 &0.614& 12.6\\
        M 30& -2.3& 12.1 &0.117& 8547 &0.170& 13.0\\
        M 92& -2.3& 12.6 &0.108& 9259 &0.072& 12.5\\
        \hline
    \end{tabular}
    \caption{Globular cluster parameters}
    \label{tab:GC}
\end{table}

\subsection{Comparison with the three ancient stars HD 84937, HD 132475, and HD 140283}
We made an additional sanity check by comparing our results with the three very old stars HD 84937, HD 132475, and HD 140283. They were studied by \cite{VandenBerg2014}. HD 140283 was studied also by \cite{Bond2013}. Both studies discovered that these stars are old and close to the age of the Universe. We used our age determination technique to derive their age. We used photometry and parallaxes from Gaia DR3, metallicity from the SIMBAD catalog\footnote{\url{http://simbad.u-strasbg.fr/simbad/sim-fbasic}}, and reddening from \cite{Lallement_2018}. All parameters are listed in Tab.\ref{tab:3 stars parameters}. The resulting ages are shown in the Tab.\ref{tab:3 stars age}. We can see that our results are close to the previously derived ages and they coincide within the uncertainties. 

\begin{table}[]
    \centering
    \begin{tabular}{lccc}
    \hline
    \hline
        parameter&HD 84937&HD 132475&HD 140283\\
        \hline
        G, mag&8.207&8.391&7.036\\
        G$_{BP}$, mag&8.423&8.692&7.321\\
        G$_{RP}$, mag&7.817&7.898&6.562\\
        $\pi$, mas&13.498&10.671&16.267\\
        $\Delta\pi$, mas&0.044&0.025&0.026\\
        $[Fe/H]$, dex& -2.0& -1.4& -2.4\\
        $[\alpha/Fe]$, dex&0.38&0.45&0.26\\ 
        A$_{V}$, mag&0.009&0.047&0.006\\
        $\Delta$A$_{V}$, mag&0.047&0.056&0.053\\
        \hline
    \end{tabular}
    \caption{\cite{VandenBerg2014} star parameters}
    \label{tab:3 stars parameters}
\end{table}

\begin{table}[]
    \centering
    \begin{tabular}{llcccc}
    \hline
    \hline
        star&source&\multicolumn{2}{c}{BaSTI}&\multicolumn{2}{c}{Padova}\\
        &&age&error&age&error\\
        \hline
        HD 84937&G vs. G$_{BP}$-G&13.18&$\pm$1.40&12.03&$\pm$1.41\\
        &G vs. G-G$_{RP}$&14.8&$\pm$1.05&13.96&$\pm$1.31\\
        &G vs. G$_{BP}$-G$_{RP}$&14.21&$\pm$1.28&13.1&$\pm$1.22\\
        &average&14.06&$\pm$1.51&13.03&$\pm$1.79\\
        \hline
        &\cite{VandenBerg2014}&12.08&$\pm$0.14&&\\
        \hline
        HD 132475&G vs. G$_{BP}$-G&13.28&$\pm$1.27&13.16&$\pm$1.14\\
        &G vs. G-G$_{RP}$&14.16&$\pm$1.27&14.09&$\pm$1.32\\
        &G vs. G$_{BP}$-G$_{RP}$&13.68&$\pm$1.14&13.64&$\pm$1.18\\
        &average&13.71&$\pm$0.81&13.63&$\pm$0.86\\
        \hline
        &\cite{VandenBerg2014}&12.56&$\pm$0.46&&\\
        \hline
        HD 140283&G vs. G$_{BP}$-G&14.02&$\pm$1.27&14.13&$\pm$1.3\\
        &G vs. G-G$_{RP}$&14.7&$\pm$1.17&15.24&$\pm$1.56\\
        &G vs. G$_{BP}$-G$_{RP}$&14.51&$\pm$1.29&14.71&$\pm$1.44\\
        &average&14.41&$\pm$0.65&14.69&$\pm$1.03\\
        \hline
        &\cite{VandenBerg2014}&14.27&$\pm$0.38&&\\
        &\cite{Bond2013}&14.46&$\pm$0.31&&\\
        \hline
    \end{tabular}
    \caption{Age derived for \cite{VandenBerg2014} stars.}
    \label{tab:3 stars age}
\end{table}

\subsection{Age averaging}
Finally, we calculated the arithmetic and weighted average age for all the stars (Tab.\ref{tab:average age}). The average age is very close to the age of the Universe that means that most of the stars under consideration were born recently after the Big Bang. The provided uncertainties take into account only the formal internal errors. But obviously we may have additional error's sources coming, for example, from isochrone's models: convection treatment, element diffusion (settlement of heavy elements and consequent wrong global metallicity derived from the actual stellar atmosphere \cite{Bonfanti_2015}) and temperature-color transformations. There are also subtle effects due to overabundance of oxygen on the "shape" of the TO-point (\cite{VandenBerg2014}) and impact in opacity from C, Ca, Mg (Sec. \ref{chemical analysis}). Another important external error could be the gas/dust ratio implied in the reddening derivation of the \cite{Lallement_2018} map based on gas. We may reasonably suppose that they account for about 0.5 Gyr (this corresponds to an Av error of $\pm$ 0.05 mag at the MS turnoff). The comparison between the two different sets of isochrones we use and our age determination of the three reference globular clusters suggest that 0.5 Gyr is a good estimation for our systematic error.

\begin{table}[]
    \centering
    \begin{tabular}{llcccc}
    \hline
    \hline
        isochrone&filter&average age&error&weighted average age&error\\
        &&Gyr&Gyr&Gyr&Gyr\\
        \hline
        Padova & G vs. BP-G & 13.6 & $\pm$1.3 & 13.7 & $\pm$1.1\\
        & G vs. G-RP & 13.9 & $\pm$1.4 & 14.1 & $\pm$1.2\\
        & G vs. BP-RP & 13.8 & $\pm$1.3 & 13.9 & $\pm$1.1\\
        & average & 13.7 & $\pm$0.5 & 13.9 & $\pm$0.5\\
        \hline
        BaSTI& G vs. BP-G & 13.4 & $\pm$0.7 & 13.5 & $\pm$0.7\\
        & G vs. G-RP & 13.6 & $\pm$0.7 & 13.9 & $\pm$0.7\\
        & G vs. BP-RP & 13.5 & $\pm$0.8 & 13.7 & $\pm$0.7\\
        & average & 13.5 & $\pm$0.4 & 13.7 & $\pm$0.4\\
        \hline
        Padova+BaSTI & average & 13.8 & $\pm$0.3 & 14.1 & $\pm$0.3\\
        \hline
    \end{tabular}
    \caption{Average age of our stars}
    \label{tab:average age}
\end{table}

\subsection{Age metallicity relation}
\begin{figure}
\centering
\includegraphics[scale=0.5]{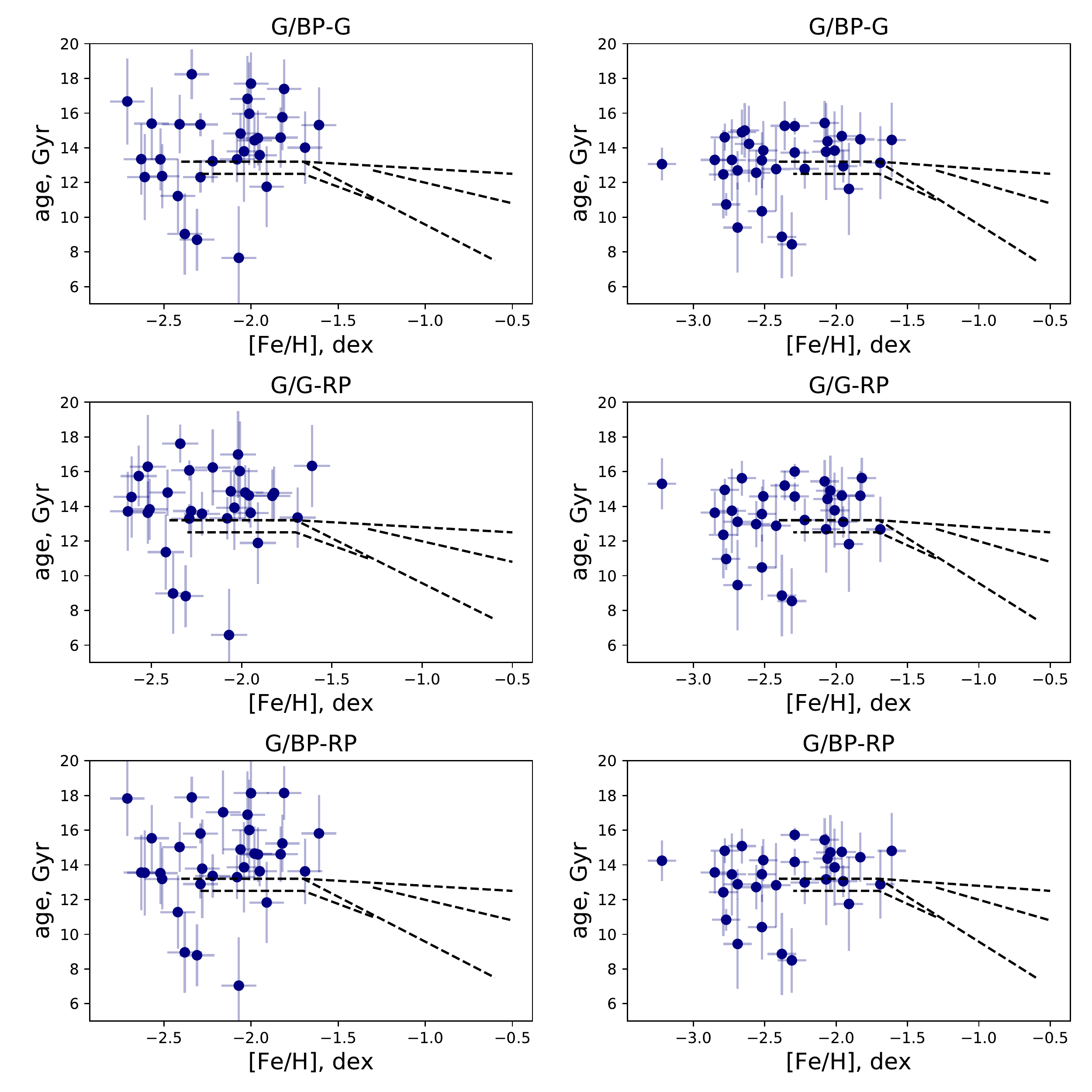}
\caption{Age metallicity relation for ages from G vs. G$_{BP}$-G, G vs. G-G$_{RP}$, G vs. G$_{BP}$-G$_{RP}$ diagrams. \textit{First column}: ages are from Padova isochrones, \textit{second column}: from BaSTI.}
\label{fig:AMR all}
\end{figure}

\begin{figure}
\centering
\includegraphics[scale=0.5]{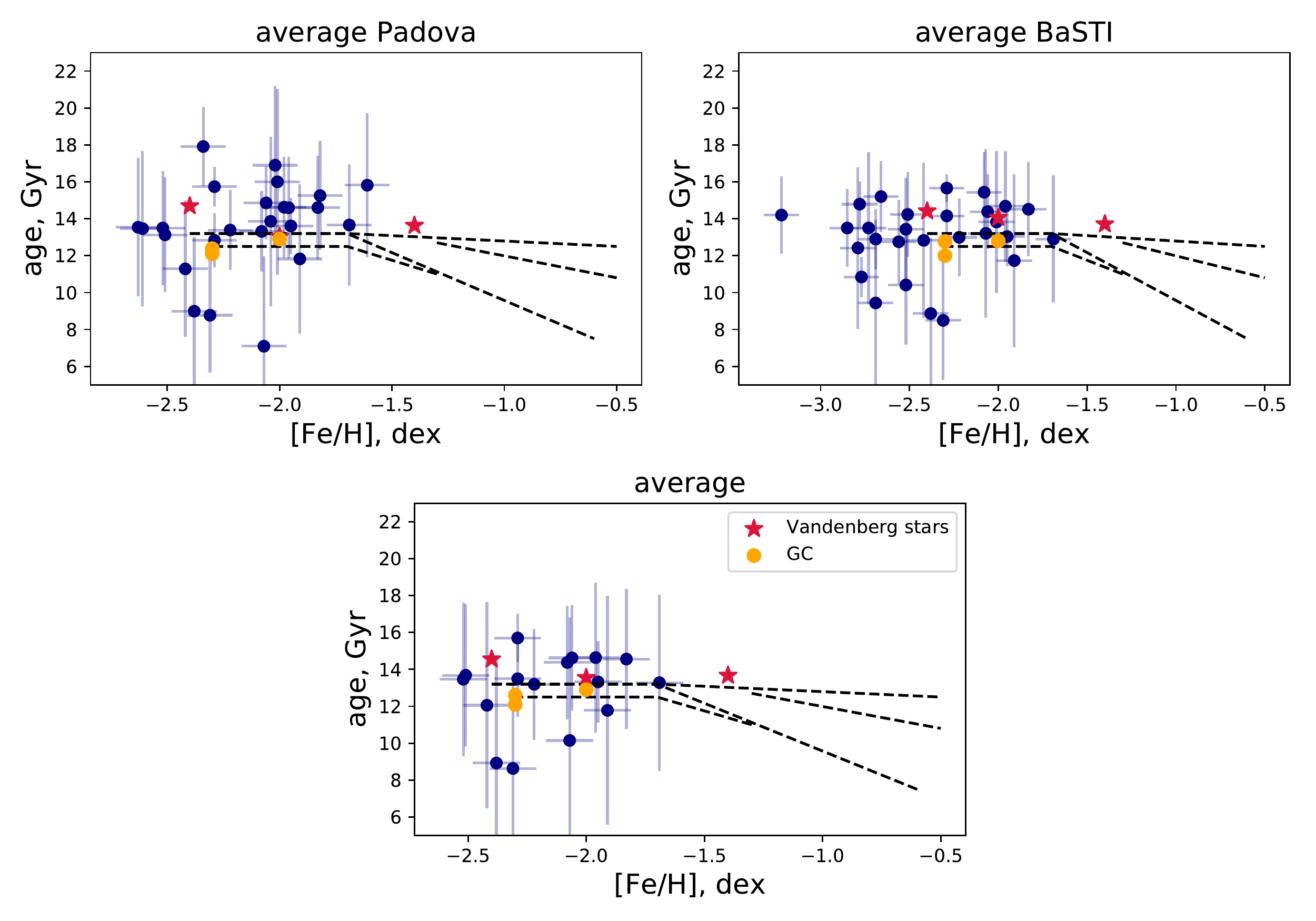}
\caption{Average age metallicity relation for Padova (\textit{left top}), BaSTI (\textit{right top}) and Padova+BASTI averaged together (\textit{bottom}). Red stars are \cite{VandenBerg2014} targets. Yellow dotes are three GC: NGC 6397, M 30, M92.}
\label{fig:AMR average}
\end{figure}

The age metallicity relation (AMR) from G vs. G$_{BP}$-G, G vs. G-G$_{RP}$, G vs. G$_{BP}$-G$_{RP}$ diagrams is presented in Fig.\ref{fig:AMR all} and for average age in Fig.\ref{fig:AMR average} together with three ancient stars from \cite{VandenBerg2014} and three metal-poor GC: NGC 6397, M30, M92. The first column show result from Padova isochrones and the second column is from BaSTI. For the comparison we took the Milky Way Globular cluster AMR from \cite{Dotter_2011} (in paper Fig.10) and  \cite{Cohen_2021} (in paper Fig.4) (dashed lines in Fig.\ref{fig:AMR all}, \ref{fig:AMR average}).
We can see that our stars extend the AMR to the lower metallicity side. On average metal-poor stars under investigation are older by about 0.8 Gyr than the main trends of the literature. And it is consistent with our results in Sec. \ref{comparison with GCs, where we showed that GCs on average show younger ages compared with single metal-poor stars}. The average age for data set under consideration is $13.7 \pm 0.4$ Gyr (BaSTI, 28 stars), $13.9 \pm 0.5$ Gyr (Padova, 17) and $14.1 \pm 0.3$ Gyr (from BaSTI and Padova, 17 stars). Moreover for metallicity between -2.7 -- -2.0 dex we have a minor population with relatively younger ages around 8 -- 10 Gyr. It can be an indication of two different populations or two epochs of star formation.

\section{Conclusion}
In this study, we derived the age of 28 metal-poor stars (Tab.\ref{tab:age}). The average age from Padova isochrones is $13.9 \pm 0.5$ Gyr, from BaSTI - $13.7 \pm 0.4$ Gyr and for star with both determination it is of $14.1 \pm 0.3$ Gyr (internal errors only). Our estimate for the additional systematic error is around 0.5 Gyr.

Age was derived by automatic isochrone CMD fitting using Gaia DR3 photometry and the most updated parameters for distance and reddening. We considered also B,V photometry. Distance was chosen among four sources: Gaia DR3 (\cite{GaiaEDR32021}), Gaia DR3 corrected by \cite{Lindegren2021},  \cite{Bailer-Jones2021} and \cite{Queiroz_2019} (\texttt{StarHorse}). Gaia DR3 parallaxes showed the best results in terms of CMD dispersion. We ascribe this result to the relative brightness of our stars. The best reddening estimate was obtained by combining two reddening maps \cite{Green_2018} and \cite{Lallement_2018} extended by \cite{Montalto_2021}.

We used two sets of isochrones, Padova and BaSTI. Padova isochrones produce on average 0.5 Gyr older age than BaSTI isochrones. 
Age determination can depend also on the detailed chemical composition of the stars. The enhancement of $\alpha$-elements, C, N, O, and Ne can significantly affects the opacity of metal-poor stars and changes the continuum emission. The strongest effect is in the Johnson B filter (Fig.\ref{fig:Spectra comparison}). The effect is marginally present also in Gaia filters but, due to their width, it is much less prominent. We suggest that, in order to improve the age determination, specific isochrones should be computed for the chemical composition of each star.

In order to minimise the chemical composition effect in this work we used 3 different filter combinations of Gaia filters: G vs. G$_{BP}$-G, G vs. G-G$_{RP}$, G vs. G$_{BP}$-G$_{RP}$. We found that the age difference from one combination to the other is about 0.2 Gyr or less, smaller than the total uncertainty of the age determination method (more than 0.5 Gyr).

We checked our results against the three most metal-poor GC: NGC 6397, M 30, M 92. GC's ages derived by our automatic isochrone fitting technique are in good agreement with ages derived by other authors (\citet{Correnti_2018} (NGC 6397), \citet{Kains_2013} (M 30), \citet{VandenBerg_2016} (M 92)). Moreover, our set of very metal-poor stars on average is older than the most metal-poor GC by about 1 Gyr.

Additionally, we compared our results with the ages of three nearby ancient halo sub-giants (\cite{VandenBerg2014}). We found that our ages coincide within the uncertainties with the \cite{VandenBerg2014} and \cite{Bond2013} ages. Results are summarized in Tab.\ref{tab:3 stars age}.

Finally, we studied the age metallicity relation in its very metal-poor tail. The trend is almost horizontal, but our stars are on average older than mean locus found by other authors (\cite{Dotter_2011}, \cite{Cohen_2021}) by about 0.8 Gyr. The interesting fact is that we found a group of very metal-poor stars with significantly young ages of 8 -- 10 Gyr. The age of these stars can be a signature of two different populations or two epochs of star formation.

Our future plans include:
\begin{itemize}
    \item[-] Extending our data set for the new available samples of metal-poor and very metal-poor stars (\cite{Li_2022}, \cite{Xu_2022}, \cite{Lucey_2022}).
    \item[-] Assessing the origin of these stars through the detailed analysis of their kinematics and chemical composition.
\end{itemize}

\begin{acknowledgments}
The comments of an anonymous referee have been much appreciated.
AP acknowledges the Ulisse program of Padova University which allowed her to spend a period at Concepcion University, where part of this works has been done. SV gratefully acknowledges the support provided by Fondecyt regular n. 1220264 and by the ANID BASAL projects ACE210002 and FB210003.
\end{acknowledgments}

\bibliography{sample631}{}
\bibliographystyle{aasjournal}



\end{document}